\documentclass[12pt]{article}

\usepackage{graphicx}
\usepackage{enumerate}
\usepackage{natbib}
\usepackage{url} 
\RequirePackage{amsthm,amsmath,amsfonts,amssymb}

\usepackage{algorithm}
\usepackage{algpseudocode}
\usepackage{subcaption}
\usepackage{setspace}

\usepackage{multirow}

\RequirePackage{graphicx}

\newcommand{\blind}{1}

\newtheorem{theorem}{Theorem} 
\newtheorem{lemma}{Lemma} 
\newtheorem{corollary}{Corollary} 
\newtheorem{propo}{Proposition} 
\newtheorem{remark}{Remark} 
\newtheorem{ass}{Assumption}
\newtheorem{definition}{Definition}

\usepackage{tikz}
\usepackage{booktabs}
\usetikzlibrary{positioning, arrows.meta}
\usepackage{enumitem}

\usepackage{tikz-cd}
\usepackage{graphicx}
\usepackage{subfig}
\usepackage{pgfplots}
\usepackage{adjustbox}
\usepackage{array}
\usepackage{pgfplotstable}
\usepackage{caption}
\usepackage{subcaption}
\usepackage{booktabs}
\pgfplotsset{compat=1.17}
\newcommand{\approxtext}[1]{\ensuremath{\stackrel{\text{#1}}{=}}}

\newcommand{\alg}{\mathcal{A}}
\newcommand{\Xcal}{\mathcal{X}}
\newcommand{\R}{\mathbb{R}}

\begin{document}
	
	 \newcommand{\myTikzPicture}[5]{
		\begin{tikzpicture}
			\begin{axis}[
				xlabel=$l$, ylabel=$u$,
				xmin=0, xmax=10, ymin=0, ymax=10,
				axis lines=middle,
				axis line style={-latex},
				width=4cm, height=4cm,
				xtick=\empty, ytick=\empty,
				xticklabels={,,}, yticklabels={,,},
				enlargelimits={abs=0.5},
				]
				\addplot[mark=*, only marks, color=#3] coordinates {(1.5,4.5)};
				\addplot[mark=*, only marks, color=#4] coordinates {(3,7.5)};
				\addplot[mark=*, only marks, color=#5] coordinates {(6.5,8.5)};
				\fill[gray!40] (axis cs:0,10) rectangle (axis cs:#1,#2); 
				\addplot[dashed, thick, domain=0:10] {x};
			\end{axis}
		\end{tikzpicture}
	}

\def\spacingset#1{\renewcommand{\baselinestretch}%
{#1}\small\normalsize} \spacingset{1}


\if1\blind
{
  \title{\bf Uncertainty quantification for intervals}
  \author{Carlos García Meixide\thanks{
    meixide@berkeley.edu}\hspace{.2cm}\\
    UC Berkeley - ICMAT\\
    and \\
    Michael R. Kosorok \\
    UNC Chapel Hill \\
    and \\
    Marcos Matabuena \\
    Harvard University}
  \maketitle
} \fi

\if0\blind
{
  \bigskip
  \bigskip
  \bigskip
  \begin{center}
    {\LARGE\bf Uncertainty quantification for intervals}
\end{center}
  \medskip
} \fi

\bigskip
\begin{abstract}
Data following an interval structure are increasingly prevalent in many scientific applications. In medicine, clinical events are often monitored between two clinical visits, making the exact time of the event unknown and generating outcomes with a range format. As interest in automating healthcare decisions grows, uncertainty quantification via predictive regions becomes essential for developing reliable and trustworthy predictive algorithms. However, the statistical literature currently lacks a general methodology for interval targets, especially when these outcomes are incomplete due to censoring. We propose an uncertainty quantification algorithm for interval responses and establish its theoretical properties using empirical process arguments based on a newly developed class of functions specifically designed for these interval data structures. Although this paper primarily focuses on deriving predictive regions for interval-censored data, the approach can also be applied to other statistical modeling tasks, such as goodness-of-fit assessments. Finally, the applicability of the method is demonstrated through simulations, showing up to a 60\% improvement in conditional coverage. Our new algorithm is also applied to various biomedical contexts, including two clinical examples: i) sleep duration and its association with cardiovascular diseases, and ii) survival time in relation to physical activity levels.

\end{abstract}

\noindent%
\vfill

\newpage
\spacingset{1.9} 

\section{Introduction}
Survival analysis constitutes core statistical methodology in biomedical research and precision medicine \citep{kosorok2019precision}, providing methods to quantify the lifetime of patients under different conditions and treatments. In these fields, survival models are the primary tools for establishing the effectiveness of drugs and clinical interventions concerning the occurrence of specific events, as is the case in clinical trials \citep{fleming2000survival}. 

\begin{figure}[h!]
	\centering
	\includegraphics[width=0.5\textwidth]{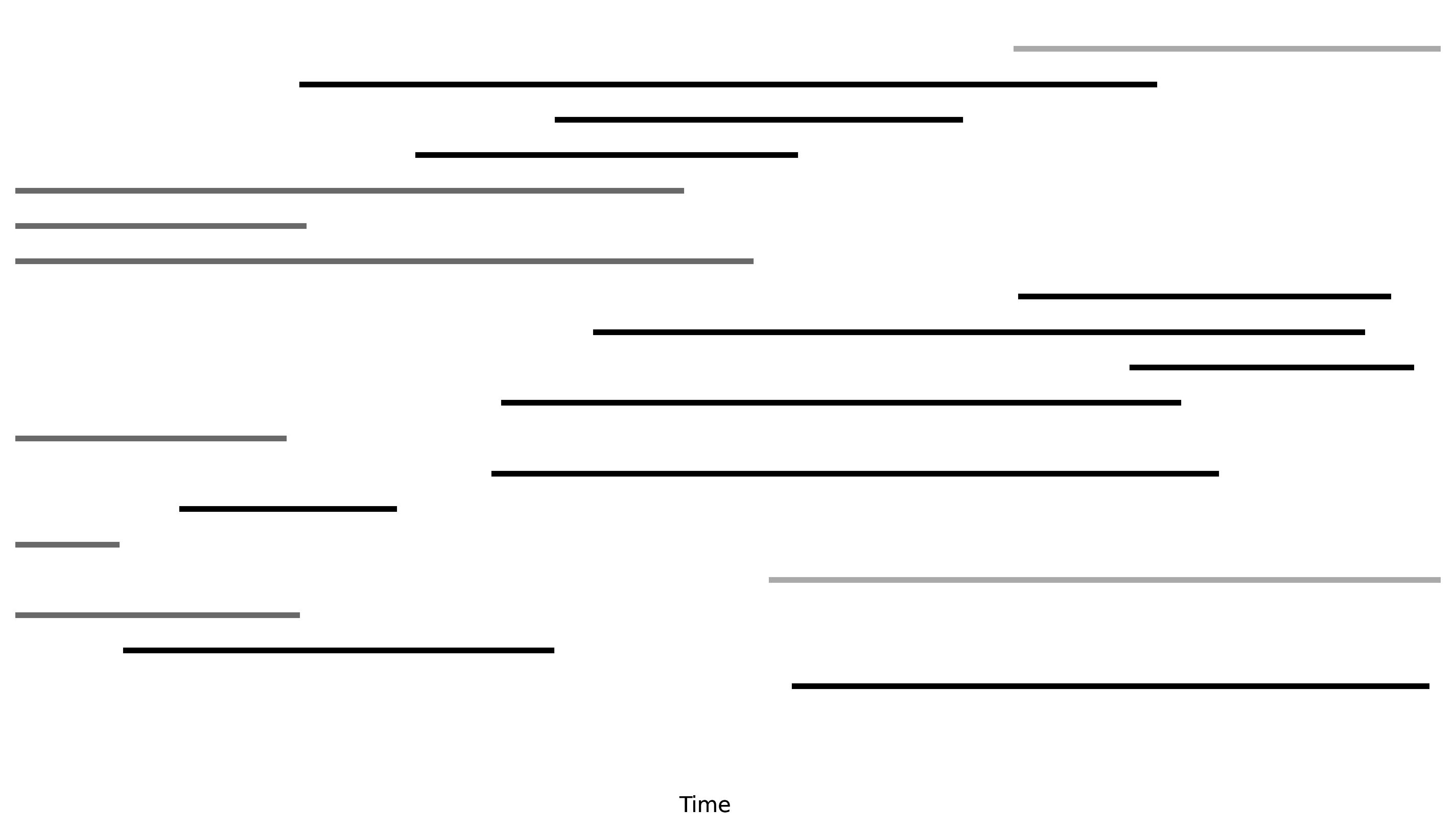}
	\caption{Illustration of a target variable with an interval structure. Each segment represents a patient. The borders of the black lines indicate medical checkpoints that bound the true, unknown time-to-event. The light grey segments represent right-censored patients, while the dark grey segments represent left-censored patients.}
	\label{fig:int}
\end{figure}

Traditionally, practitioners have focused on right-censored outcomes, which constitute the conventional setting in survival analysis \citep{kaplan1958nonparametric}. However, in other clinical applications, such as digital health, outcomes often possess an interval structure.  For instance, with electronic records \citep{hyun2017flexible}, the time to a clinical event may occur between two medical visits, as seen in the case of diabetes mellitus or other chronic diseases. Another example of an interval setup is found in clinical settings related to sleep studies. Here, patients typically record the time they go to bed and the time they wake up; however, the precise moment they actually fall asleep is not documented. This results in an interval of time during which the event (falling asleep) occurs, but the exact time within this interval remains unknown. Similarly, with wearables, devices are not worn continuously, and some defined clinical events may occur between two wear periods. Given that patient monitoring is often quasi-continuous, it is imperative to develop new methodologies for handling these interval-structured outcomes. A comprehensive illustration is depicted in Figure \ref{fig:int}.

Quantifying the uncertainty of predictive outcomes is one of the most crucial modeling tasks in statistical sciences, particularly in clinical applications within personalized medicine. In clinical studies, the response of patients over time exhibits individual and heterogeneous behavior. Therefore, when predicting clinical outcomes, there is considerable variability in patient responses. Reporting only point estimates, such as the conditional mean, provides partial and imperfect knowledge of the phenomenon under study and can lead to misleading clinical conclusions regarding the effects of drugs and interventions \citep{candesrssb}. Consequently, measures of uncertainty, such as predictive regions, must be included in analyses. Although the construction of covariate-dependent predictive regions has received significant attention throughout the historical development of statistics \citep{wilks1941determination,vovk2005algorithmic}, the literature on survival analysis and censored outcomes remains more limited.
With the proliferation of data-driven systems across various scientific and engineering fields, quantifying uncertainty has become a central research topic within the statistical and machine learning communities.

A particularly prolific framework that has emerged in recent years is conformal inference \citep{vovk2005algorithmic}, which encompasses a general family of methods. Given a random sample $\mathcal{D}_{N} = \{(X_i, T_i) \in \mathcal{X} \times \mathcal{T} : 1 \leq i \leq N \}$, assumed to be at least exchangeable, and a new observation $(X_{N+1}, T_{N+1}) \in \mathcal{X} \times \mathcal{T}$ with the same exchangeability property relative to $\mathcal{D}_{N}$, it is possible to consider a predictive region $\widehat{\mathcal{C}}_{1-\alpha}(\cdot) \subset \mathcal{T}$ based on $\mathcal{D}_{N}$ such that the following non-asymptotic property is satisfied: $\mathbb{P}(T_{N+1} \in \widehat{\mathcal{C}}_{1-\alpha}(X_{N+1})) \geq 1 - \alpha.$

\noindent Here, $\mathbb{P}$ represents the joint probability law over the entire dataset $\mathcal{D}_N$ and the new data point $(X_{N+1}, T_{N+1})$. Other methods in the data analysis literature that address similar scientific problems include Bayesian methods, which require the subjective specification of priors, asymptotic approximations based on the central limit theorem, and, more broadly, bootstrapping techniques \citep{stine1985bootstrap}.

The key characteristic of the various methods mentioned is that they operate on scalar, multivariate, and functional random variables $T \in \mathcal{T}$ (see review \cite{lugosi2024uncertainty}). However, in several scientific applications, the only available information is a random interval $(L, U)$, where $0 \leq L \leq U \leq \infty$ and $T \in (L, U) \subset \mathcal{T}$ $\mathbb{P}$-a.s. This setup encompasses right-censoring when $U = \infty$ and left-censoring when $L = 0$.

General methodologies for providing predictive regions of an interval-censored target typically focus on estimating the conditional distribution \citep{huang1996efficient}, but they may result in poorly calibrated predictive intervals and a higher dependence on the chosen model. The non-parametric estimation of survival functions with interval-censored data is based on \citep{turnbull1976empirical}'s algorithms, which can be derived using a self-consistency argument \citep{efron1967two}. Efron also proved that this approach yields the nonparametric maximum likelihood estimator (NPMLE). However, significant issues highlighted in the literature include the lack of uniqueness when the log-likelihood is not strictly concave, and the fact that consistency results are typically available only under the assumption of a fixed number of inspection times \citep{gentleman}. \citep{Maathuis2008-xg} deduce that the NPMLE for the joint distribution function of an interval-censored survival time and a continuous mark variable is generally not consistent. Furthermore, the self-consistent approach may not be consistent even when the inspection times are limited to finitely many values, as demonstrated by a counterexample in \citep{yu2000consistency}. Additionally, traditional methodologies in computational statistics, such as the bootstrap, are known to be inconsistent when using the NPMLE as the reference measure in the presence of interval censoring \citep{sen}. Due to these challenges, practitioners often resort to parametric and semi-parametric approaches for modeling interval-censored data, as these methods are numerically more stable and yield consistent solutions from both optimization and statistical perspectives \citep{meixidesadm}.

To address this gap in the literature, the goal of this paper is to introduce a new uncertainty quantification algorithm, \texttt{uncervals}. Our proposal is inspired by a combination of conformal prediction and bootstrap methods \citep{sen}, following the results in \cite{politis2015model}, which suggest that bootstrap improves conditional coverage. This serves as the rationale for combining conformal prediction with bootstrap in our approach: to enhance finite-sample performance in terms of marginal and conditional coverage. See the simulations in Section \ref{sim:cond}.

Our approach achieves significantly better calibration, as confirmed by our simulation studies, which are consistent with previous work (see, for example, \cite{zhang2022regression}). The new method could be applied to develop novel algorithms for interval-censored data, including variable selection methods and goodness-of-fit tests.
To derive the theoretical properties of \texttt{uncervals}, we introduce a new class of functions that leverages empirical process theory and turns out to be highly relevant in clinical applications for precision medicine and digital health, where outcomes often exhibit an interval structure.

The main contributions of this paper are summarized below:

\begin{enumerate}
	\item \textbf{Practical implications:} We introduce a comprehensive uncertainty quantification algorithm, \texttt{uncervals}, which is compatible with off-the-shelf regression models tailored for interval-censored data. 
	\item \textbf{Theoretical support:} To derive the formal guarantees of \texttt{uncervals}, we introduce a new class of functions that inherently capture the statistical nature of interval data. This enables the use of empirical process theory to establish consistency and convergence rates.
	\item \textbf{Biomedical applications:} We illustrate the potential of \texttt{uncervals} through relevant cases involving interval-censored data. We compare our approach to naive interval-censored adaptations of algorithms originally crafted for right-censoring. For these purposes, we examine the performance of our proposal across different scenarios and semi-parametric models.
\end{enumerate}

\subsection{Paper Structure}
The structure of the paper is defined as follows. Section \ref{sec:related} describes the key literature on uncertainty quantification and conformal prediction for right-censored data, providing a natural reference context to depart to interval censoring. Section \ref{sec:algorithms} introduces our algorithms for uncertainty quantification with interval-censored data and their theoretical analysis (Section \ref{sec:theory}). Section \ref{sec:simulation} presents a simulation analysis to examine the theoretical properties of \texttt{uncervals}. Section \ref{sec:realresults} introduces two examples to illustrate the potential of the methodology in biomedical applications. Finally, Section \ref{sec:discuss} discusses the contributions, limitations and future research directions of our framework.

\subsection{Related work}\label{sec:related}

We present a literature review to derive prediction regions in survival analysis based on conformal prediction techniques. Notably, our proposal is designed for interval-censored data, whereas the existing literature focuses on right-censored outcomes. The right-censored case can be considered a special case of interval-censored data; however, despite this, our method transcends being a mere extension of right-censored approaches, which do not have the potential to operate effectively for interval-censored data.

Survival analysis traditionally aims to infer the probability that a patient will survive beyond a specified time. This is often complicated by the presence of censoring, where for some patients, survival times are only known to exceed a certain value due to limitations such as the end of a study period. Machine learning methods are introduced as promising tools for handling such complex data without relying on strong modeling assumptions. However, these methods face challenges in quantifying uncertainty, which is critical for making reliable predictions in high-stakes situations. To overcome this problem, \cite{candesrssb} present a methodology to construct prediction intervals for survival times under the Type-I censoring setting, which assumes that the censoring time is observed for every unit, while the outcome is only observed for uncensored units. This is more restrictive than traditional right-censoring, where the censoring time is not available for censored units. This new lens on the problem is particularly relevant in modern healthcare applications, such as the COVID-19 pandemic, where accurate prediction of survival times is crucial for resource allocation and decision-making in public health crises. Refer to the supplement for a comprehensive review of the statistical literature on conformal prediction. 

The type of predictive region \cite{candesrssb} propose is called the lower predictive bound (LPB), which can be seen as a one-sided tolerance region. The LPB serves as a conservative estimate of the survival time, offering a critical tool for high-stakes decision-making by ensuring that the true survival time falls above this bound with a specified probability. We say an LPB $\hat{L}_{1-\alpha}(\cdot)$ is calibrated if $ \mathbb{P}(T_{N+1} \geq \widehat{L}_{1-\alpha}(X_{N+1})) \geq 1-\alpha,$ where $\alpha$ is a pre-specified level, and the probability is computed over both $\mathcal{D}_N$ and a future unit $(X_{N+1}, T_{N+1})$ that is independent of $\mathcal{D}_{N}$. This can be seen as a predictive interval whose upper border is $+\infty$, serving as a conservative assessment of survival.

\begin{figure}[t!]
	\centering
	\begin{subfigure}[t]{0.5\textwidth}
		\centering
	\includegraphics[width=\textwidth]{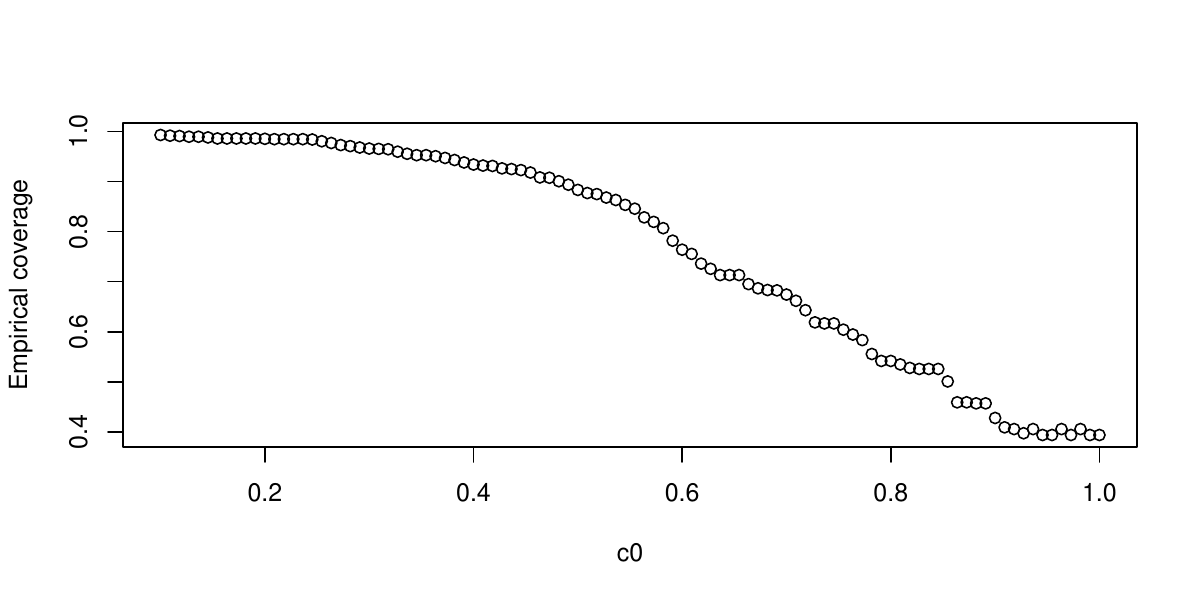}
			\caption{Evolution of empirical coverage with the threshold $c_0$ used in \cite{candesrssb}. The desirable outcome would be a horizontal line at 0.9. Setup details in \ref{evalcfs}.}
			\label{fig:c00}
		\end{subfigure}
		\hfill
		\begin{subfigure}[t]{0.45\textwidth}
			\centering
			\includegraphics[width=\textwidth]{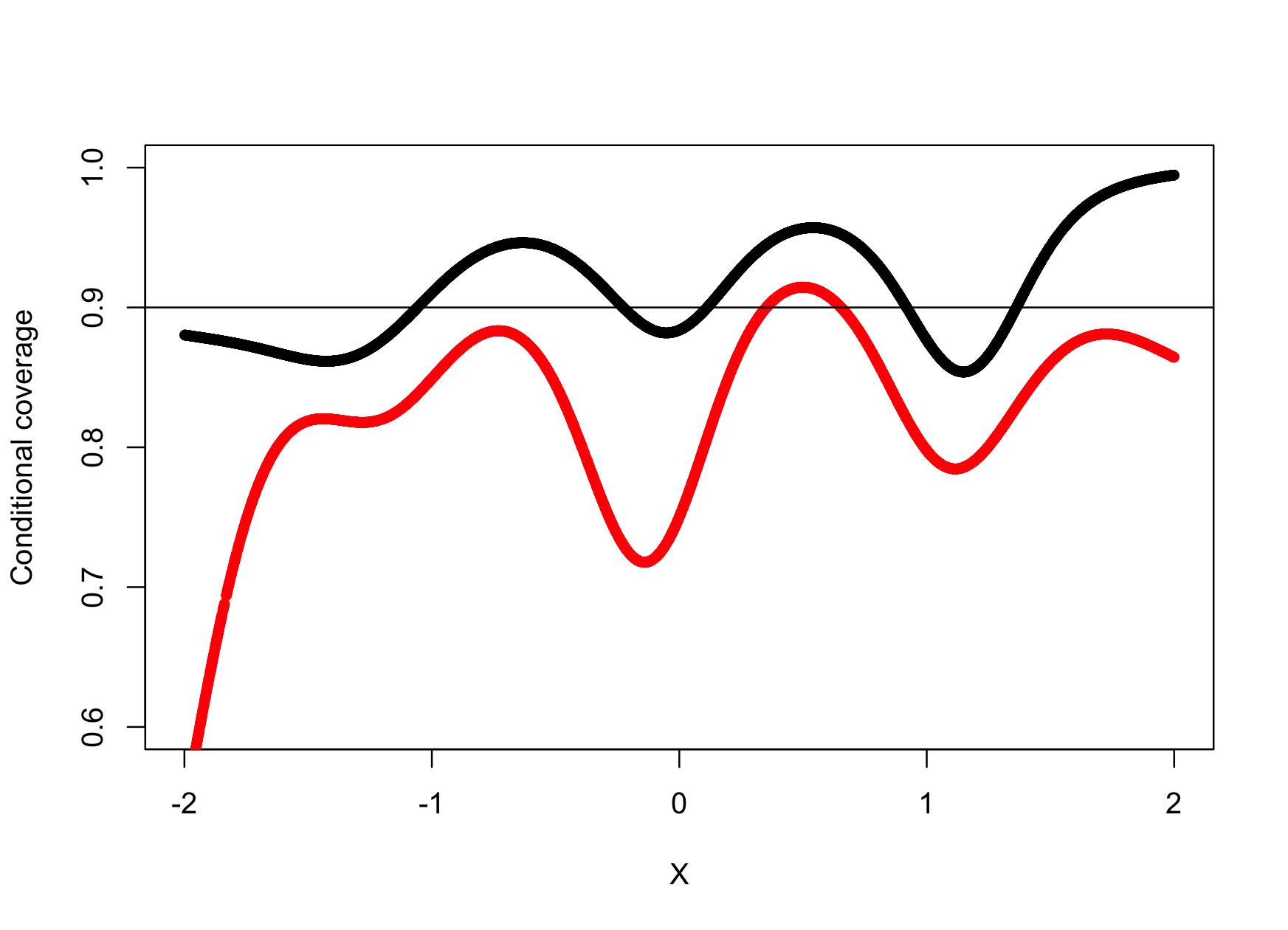}
			\caption{Red: conditional coverage achieved by naively taking the $\alpha$-th quantile of a distributional regression algorithm. Black: our approach. Details in Section \ref{sim:cond}  }
			\label{fig:condcov}
		\end{subfigure}
		\caption{Failure of uncertainty quantification approaches for interval data}
	\end{figure}

Let $\tilde{T}_i = \min(T_i, C_i)$. Upon making the following elementary observation:
\begin{equation}\label{basic}
	\tilde{T}_i \leq T_i \Longrightarrow \textrm{ any calibrated LPB on } \tilde{T}_i \textrm{ is also a calibrated LPB on } T_i
\end{equation}

\noindent a naive approach to construct distribution-free LPBs in right-censored scenarios immediately arises: just apply conformal prediction over the censored target $\tilde{T}$ in a one-sided fashion. However, this approach is overconservative in the sense that it will lead to an excessively low $\hat{L}_{1-\alpha}(\cdot)$. \cite{candesrssb} observe that this phenomenon is related to censoring times being overall smaller than survival times. This led them to apply conformal inference on subpopulations with larger censoring times while correcting for the covariate shift this selection produces applying a one-sided version of weighted conformal inference \citep{weighted}. It can even be shown that if $(T, X) \perp C$, then the covariate shift vanishes and the technique reduces to applying conformal inference on $\left(X_i, T_i \wedge c_0\right)_{C_i \geq c_0}$. This has two main problems:
\begin{itemize}
	\item In general, there is no access to all the censoring times in practical situations. This would involve, for instance, knowing when a person would quit from a clinical trial even though this individual has experienced the event of death.
	\item Their approach depends on a hyperparameter $c_0$ that selects a subpopulation with censoring times bigger than $c_0$. Figure \ref{fig:c00} serves as an illustration of how sensitive the approach by \cite{candesrssb} is to the hyperparameter $c_0$.
\end{itemize}

 {Generalizing the existing non-asymptotic guarantees for right-censored data to new ones that account for interval-censoring is highly non-trivial. This difficulty arises in part from the fact that, under interval-censoring, survival times are never observed—we only have access to $(L, U)$, which almost surely contains the true survival time $T$. We begin our arguments by building upon an observation analogous to (\ref{basic}):}

\begin{equation}\label{gen}
	L_i \leq T_i \Longrightarrow \textrm{ any calibrated LPB on } L_i \textrm{ is also a calibrated LPB on } T_i
\end{equation}

 The first immediate issue of this approach is that it would lead to even less tight LPBs. However, there is a clear benefit from this, which is the well-known finite sample size coverage guarantee inherent to conformal inference, as stated in Theorem \ref{finitethm}. 
 
  {An interval whose lower and upper borders are respectively the $\frac{\alpha}{2}$ and $1-\frac{\alpha}{2}$ quantiles of $T$ given $X=x$ is conditionally valid. This would be brought into practice by plugging consistent estimators of such quantiles, recoverable from the conditional survival function estimated by, for instance, interval-censored recursive forests \citep{icrf}.  }

 {Other recent notable work include \cite{sesia2024}, which proposes a conformal inference method for constructing lower prediction bounds for survival times from right-censored data by imputing unobserved censoring times and then accounting for covariate shift \citep{weighted}. \cite{farina2025} introduce two new methods for right-censored outcomes based on inverse-probability-of-censoring weighting and augmented inverse-probability-of-censoring weighting.}

 {These ideas, originally designed for right-censoring, do not easily transfer to the context of interval-censored outcomes. This due to the fact that in right-censored we actually manage to observe times, while in interval-censoring point observations are never sampled. Our strategy differs both theoretically and practically, as we will see next.}

\section{Predictive regions for interval-censored targets}\label{sec:algorithms}

\subsection{Setup and notation}

Let $T$ be an a.s. positive random variable representing true event times. In our setup, there will never be access to a sample from $T$ but from $L$ and $U$, which are two neighboring checkup points quantifying the interval where $T$ is known to lie within. Let $X$ be real $p$-dimensional covariates. Our sample is $(L_1,U_1,X_1), (L_2,U_2,X_2), \ldots, (L_N,U_N,X_N) \sim (L,U,X) \textrm{ i.i.d}$,	where we assume the existence of underlying $T_1,T_2, \ldots, T_N$ respecting $L_i \leq T_i \leq U_i$ almost surely for each $1 \leq i \leq N$. The formalism we propose includes interval-, right- and left-censoring observations as special cases. We encode with $U_i= \infty$ a right-censored observation and with $L_i= 0$ a left-censored observation. Otherwise, $(L_i,U_i)$ represents an interval-censored observation.

Let $F(t, x)=P\left(T_i \leq t \mid X_i=x\right)$ denote the conditional cumulative distribution function (cdf) of $T_i$ given $X_i=x$. Throughout the paper, we will split the whole data $\{1, \ldots, N\}$ into $\mathcal{I}_1:=\{n+1, \ldots, N\}$ and $\mathcal{I}_2:=\{1, \ldots, n\}$. Because of being our data i.i.d, random splits are also possible. We assume we have obtained an estimator $\hat F_1$ of $F$ based on $\mathcal{I}_1$ and we denote by $\Lambda_i:=\hat F_1(L_i, X_i)$, $\Upsilon_i:=\hat F_1(U_i, X_i)$ the evaluations of $\hat F_1$ on observable data coming from $\mathcal{I}_2$ and indexed by $i=1,\ldots,n$ according to how it was chosen. We denote by $\Phi_i:=\hat F_1(T_i, X_i)$ the evaluations of $\hat F_1$ on the underlying survival times $T_i$ (we do not have empirical access to $\Phi_i$, but we need to postulate them). Upon fixing $\left\{(L_i, U_i, X_i)\right\}_{i\in \mathcal{I}_1}$- and therefore also $\hat F_1(\cdot,\cdot)$- then $\left\{(\Phi_i,\Lambda_i, \Upsilon_i)\right\}_{i\in \mathcal{I}_2}$ are i.i.d. because they are evaluations on an i.i.d. sample of a function that became deterministic after conditioning. We denote by ${P}$ the conditional-on-split-1 joint law of $(\Phi,\Lambda, \Upsilon)$ and by $G$ the population cdf of $\Phi_1, \ldots, \Phi_n$ conditional on $\hat F_1$, respectively. 	$\mathbb{P}$ denotes the joint probability law over $\mathcal{D}_N$ and $(X_{N+1},T_{N+1})$; $\mathbf{P}(\cdot)= \mathbb{P}(\cdot \cap \mathcal{I}_1)$ the marginal probability law of split 1; and $P(\cdot)= \mathbb{P}(\cdot | \mathcal{I}_1)$ the probability law conditional on split 1. The sample structure is as follows, where we chose split 2 to be before split 1 just for notational convencience:
$$\underbrace{1, \ldots, n-1,n}_{\text{split 2}}, 
\underbrace{n+1, \ldots, N-1, N}_{\text{split 1}}, 
\underbrace{N+1}_{\text{new}}.
$$

We present \texttt{uncervals} in Algorithm \ref{oa}, where the user is free to choose between two operating modes indexed by $e=0,*$. 

\singlespacing
\begin{algorithm}[h!]
	\caption{\texttt{uncervals}}\label{oa}
	\begin{algorithmic}[1] 
		\Require $\left\{(L_i, U_i, X_i)\right\}_{i=1}^N$, $\alpha$, $X_{N+1},e$
		\Statex
		
		\State Split $\{1, \ldots, N\}$ into $\mathcal{I}_1:=\{ n+1, \ldots, N\}$ and $\mathcal{I}_2:=\{1, \ldots, n\}$
		
		\State Obtain $\hat F_1$ ($1 - $ conditional survival function) based on $\mathcal{I}_1$ using an off-the-shelf semi-parametric model for interval-censored responses.
		
		\For{$i = 1,\ldots,n $}
		\State $j \gets \textrm{Uniform}(\{1,\ldots,n\})$
		\If{$e=0$}
		\State $\Phi^*_i \gets$ $\hat F_1(L_j, X_j)$
		\Else 
		\State  $\Phi^*_i \gets$ $\hat F_1(L_j, X_j) +( \hat F_1(U_j, X_j) - \hat F_1(L_j, X_j)) \cdot \textrm{Uniform}(0,1)$
		\EndIf
		
		\State $V_i^* \gets \psi\left({\Phi}_i^*\right)$
		\EndFor
		
		\State $\hat{Q}_{\mathcal{I}_2}^* \gets (1-\alpha)\left(1+1 /n\right)$ empirical quantile of $\left\{V_i^*\right\}$
		
		Final $(1-\alpha)$ prediction set  $\widehat{\mathcal{C}}_{1-\alpha}^{e}(X_{N+1}) \gets$  $\left\{t \geq 0: \psi(\hat{F}_1(t, X_{N+1})) \leq \hat{Q}^*_{\mathcal{I}_2}\right\}$
		
	\end{algorithmic}
\end{algorithm}
\doublespacing

In Algorithm \ref{oa},  $\psi (\cdot):= \lvert \cdot - b\rvert, b \in [0,1]$. When $F(\cdot, x)$ is a symmetric unimodal distribution with a well-defined conditional density, then $b=1/2$ is optimal in the sense that the interval outputted by Algorithm \ref{oa} has minimum length asymptotically\footnote{The fact that $b=1/2$ leads to intervals that are asymptotically similar to the oracle has nothing to do with optimal (smallest) length.}. An extension to choose $b$ optimally to ensure efficiency as in \cite{cherno} would be certainly possible, but in this article we are interested in LPBs as is natural in survival analysis applications, so that we set $b=1$ unless explicitly stated. Observe that after settng $b=1$, $$\begin{aligned}\widehat{\mathcal{C}}_{1-\alpha}^{ e}\left(X_{N+1}\right) =& \left\{t: \left|\hat{F}_1\left(t, X_{N+1}\right) - 1 \right| \leq \hat{Q}_{\mathcal{I}_2}^*\right\} \\  = &\left\{t: 1- \hat{F}_1\left(t, X_{N+1}\right) \leq \hat{Q}_{\mathcal{I}_2}^*\right\}=\left\{t:  \hat{S}_1\left(t, X_{N+1}\right) \leq \hat{Q}_{\mathcal{I}_2}^*\right\}\end{aligned}.$$	

As (empirical) survival functions are monotonically decreasing we have that, upon setting $b=1$, the previous set is indeed given by an LPB $(\widehat {L}_{1-\alpha},+\infty)$ such that: $$\widehat {L}_{1-\alpha}(X_{N+1})= \inf \{ t : {\hat{S}_1\left(t, X_{N+1}\right) \leq \hat{Q}_{\mathcal{I}_2}^*}\} $$

The operating mode $e=0$ is a generalization of \cite{candesrssb} to interval-censoring in virtue of (\ref{gen}), which enjoys finite sample size calibration guarantees but pays the price of overconservativeness; as the original approach does. There is no randomness involved at all when $e=0$.

\begin{theorem}[Finite-sample validity of \texttt{uncervals}]\label{finitethm}
	
	Set $e=0$, or equivalently, just consider the evaluations of $\hat F_1$ on left borders. Suppose that the data are i.i.d. or exchangeable and that the estimator of the conditional survival function is invariant to permutations of the data. Then $\widehat{\mathcal{C}}_{1-\alpha}^{ 0}(\cdot)$ is an LPB and 
	$$
	\mathbb{P}\left(T_{N+1} \in \hat{\mathcal{C}}_{1-\alpha}^{\text {0 }}\left(X_{N+1}\right)\right) \geq 1-\alpha .
	$$
\end{theorem}

We study in the next section the theoretical properties of the mode $e=*$, which has provable desirable asymptotic coverage.

\subsection{Statistical theory: the interval process}\label{sec:theory}
The main goals of this section are to show that \texttt{uncervals} operating with $e=*$ provides asymptotically valid prediction intervals and to derive at which rate this happens. We have introduced \texttt{uncervals} as a meta-algorithm that in practice is constituted by a symbiosis between conformal prediction and the bootstrap. However, from a theoretical perspective its foundations are rooted in a new class of functions whose properties capture the phenomenology of interval data and enable the establishment of performance guarantees through VC theory arguments. 

Upon fixing $\hat F_1$, we have transitioned to an auxiliary sample $(\Lambda_1, \Upsilon_1), (\Lambda_2, \Upsilon_2),\ldots, (\Lambda_n, \Upsilon_n)$ which is i.i.d. [or exchangeable if the original sample was so] after having fixed $\hat F_1$ and respects the constraint $0\leq \Lambda_i \leq \Phi_i \leq \Upsilon_i \leq 1 $ P-a.s. for each $1 \leq i \leq n$. Classical split distributional conformal prediction relies on the $(1-\alpha)\left(1+1 /n\right) $ empirical quantile  of $\psi (\Phi_1), \ldots , \psi (\Phi_2)$. In the presence of interval censoring, we do not have access to $\Phi_1, \ldots, \Phi_n$ and even less to the aforementioned empirical quantile. 
However, we will use the $\Phi_i$'s as auxiliary entities to support our theoretical arguments, even though they are never utilized by Algorithm \ref{oa}. 

\begin{remark}
	Note that simply $\Upsilon_i=1$ if observation $i$ is right-censored. 
	\end{remark}

We will use the following conceptual bridge to open a connection between conformal inference and empirical process theory. Suppose $\Phi_1, \ldots, \Phi_{n}$ are exchangeable random variables free from ties. The rank of the $n$-th observation $\Phi_{n}$ among $\Phi_1, \ldots, \Phi_{n}$ is uniformly distributed over the set $\{1, \ldots, n\}$. What is to say, under exchangeability we have $\mathbb{P}\left\{\frac{\textrm{rank}(\Phi_{n})}{n}>\epsilon\right\} \geq 1-\epsilon,
\textrm{ for all } \epsilon \in[0,1]$. The key connection with empirical process theory is the following observation:$$
\begin{aligned}
	\frac{\textrm{rank}(\Phi_{n})}{n} &= \frac{\#\left\{i=1, \ldots, n: \Phi_i \geq \Phi_{n}\right\}}{n}  &=\frac{1}{n}\sum_{i=1}^n  1\{\Phi_i \geq \Phi_{n}\}  &= 1- \texttt{ecdf}_{\Phi_1,\ldots,\Phi_{n}}(\Phi_{n})
\end{aligned}
$$
In this way, the the problem reduces to the study of the asymptotic properties of the empirical distribution function of $\Phi_1, \ldots, \Phi_{n}$. One of the core mechanisms of split conformal is the following fact: if original data was i.i.d. [resp. exchangeable], then conditional on split 1 scores are i.i.d [resp. exchangeable] even though the model under which $\hat F_1$ was fit had been misspecified. If we condition on $\mathcal{I}_1$, the calibration residuals built by evaluating on split 2 and the test residual are all i.i.d. [resp. exchangeable].  We will see that it is the theoretical mechanism inherent to split conformal of conditioning on a subset of data what makes bootstrap being valid. Under interval censoring, we lose access to the $\Phi_1, \ldots, \Phi_n$, but we will see that it is still possible to recover bootstrap replicates $\Phi^*_1, \ldots, \Phi^*_n$ from a very particular distribution coined by us the \textit{interval measure}.

\begin{definition}
	
	For $0 \leq t \leq 1$ given a sample $(\Lambda_1, \Upsilon_1), (\Lambda_2, \Upsilon_2),\ldots, (\Lambda_n, \Upsilon_n)$ satisfying $ 0\leq \Lambda_i \leq \Upsilon_i \leq 1 $ a.s. for each $1 \leq i \leq n$, we define  the interval measure as $$\mathbb{Q}_n(t):=\frac{1}{n} \sum_{i=1}^n\frac{1}{\Upsilon_i-\Lambda_i}1_{(\Lambda_i,\Upsilon_i)}(t)$$

\end{definition}

Having access to a hypothetical sample $\left\{(\Phi_i,\Lambda_i, \Upsilon_i)\right\}_{i=1}^n$ (which is not the case because we miss the $\Phi_i$'s), the \textit{empirical measure} is traditionally defined for $0 \leq t,\phi,\lambda,\upsilon \leq 1$  as $P_n(\phi,\lambda,\upsilon)=\frac{1}{n} \sum_{i=1}^n \delta_{\Phi_i,\Lambda_i,\Upsilon_i}(\phi,\lambda,\upsilon)
$. The first observation to be made is that, in contrast with the empirical measure, the interval measure is not composed by atoms placed on the observations but by probability boxes. 

It is easy to prove by basic integration of constant functions that the length of $[0,t]$ with respect to the interval measure is given by 
$$
\begin{aligned}
	&\int_0^t d\mathbb{Q}_n(t')  =\frac{1}{n} \sum_{i=1}^n \frac{1}{\Upsilon_i-\Lambda_i} \left[ (t-\Lambda_i) 1_{[0, t]} (\Lambda_i) 
	+  (\Upsilon_i-t) 1_{[0, t]} (\Upsilon_i) \right] &=:\mathbb{I}_n(t)
\end{aligned}$$

\noindent We call $\mathbb{I}_n(t)$ the \textit{interval distribution} for $0 \leq t\leq 1$. It is important not to miss the original sample picture:

\adjustbox{scale=1.1,center}{%
	\begin{tikzcd}
		\mathcal{I}_1\arrow{r}   & \widehat{F}_1 \arrow{d} \\
		\mathcal{I}_2\arrow{r} & \phantom{s}\mathbb{I}_n
	\end{tikzcd}
}

\noindent Now let $\mathcal{G}=\left\{g_t(l, u): t \in[0,1]\right\}$, where
$$
g_t: (l, u) \mapsto 1_{\{0 \leq l \leq u \leq 1\}} \frac{1}{u-l}\left[(t-l) 1_{[0, t \mid}(l)+(u-t) 1_{[0, t]}(u)\right] .
$$
\begin{remark}\label{equal}
	Note that for $l = u =: w$ with $ 0 \leq w \leq 1$ we have $g_t(w,w) = 1_{[0,t]}(w), 0 \leq t \leq 1 $
\end{remark}

Remark \ref{equal} asserts that when the borders of an inteval collapse into one number, the contribution of such observation to the ecdf is a step function - as happens in the complete information regime. 

Let $t \in [0,1]$. Then $\mathbb{I}_n(t)=P_ng_t$. Provided that $\mathbb{E}g_t(\Lambda,\Upsilon)= G(t)$, where $G$ is the true cumulative distribution function of the $\Phi_i's$, the interval process $\sqrt{n}\left(\mathbb{I}_n-G\right)$ coincides with the abstract empirical process indexed by the class $\mathcal{G}$: $\{\sqrt n (P_n - P)g_t,\quad  g_t \in \mathcal{G}\} $. As $\mathcal{G}$ has a finite envelope function, the map $\sqrt n (P_n - P)=: \mathbb{H}_n$ can be viewed as an element in $\ell^{\infty}(\mathcal{G})$ classically known as the \textit{empirical process}. To summarize: 

\begin{tabbing}
	\hspace{3cm} \= \kill
	
	$\mathbb{Q}_n$ \> Interval measure, $\mathbb{Q}_n(t):=\frac{1}{n} \sum_{i=1}^n\frac{1}{\Upsilon_i-\Lambda_i}1_{(\Lambda_i,\Upsilon_i)}(t)$ \\
	$\mathbb{I}_n$ \> Interval distribution, $\mathbb{I}_n(t)=\int_0^t d\mathbb{Q}_n(t')$ \\
	$\mathbb{H}_n$ \> Abstract empirical process, $\sqrt{n}(P_n - P)$ \\
	$\mathbb{H}^*_n$ \> Bootstrap empirical process, $\sqrt{n}(P^*_n - P_n)$ \\
	
\end{tabbing}

We are intereseted in boostrapping conformity scores from $\mathbb{H}_n$. To ensure that boostrap replicates asymptotically the correct distribution we need to show that i) $\mathcal{G}$ is a universal Donsker class of functions and ii) the mean of $\mathbb{H}_n$ is zero. 

\subsubsection{Donsker-universality}

The main theoretical result of this paper is the proof that the abstract empirical process converges weakly in $\ell^\infty(\mathcal{G})$, meaning that $\mathcal{G}$ is a Donsker class. This is particularly useful because the bootstrap method is always asymptotically valid for Donsker classes of functions.

\begin{theorem}\label{th_donsker} $\mathcal{G}$ is Donsker for any distribution on $(L, U)$ such that $0 \leq L \leq U \leq 1$ almost surely. Moreover, $\mathcal{G}$ has a finite envelope function. 
\end{theorem}

\noindent 
\begin{figure}
	\centering
	\begin{minipage}{0.25\textwidth}
		\myTikzPicture{0.5}{0.5}{black}{black}{black}
	\end{minipage}%
	\begin{minipage}{0.25\textwidth}
		\myTikzPicture{2}{2}{red}{black}{black}
	\end{minipage}%
	\begin{minipage}{0.25\textwidth}
		\myTikzPicture{3.5}{3.5}{red}{red}{black}
	\end{minipage}%
	\begin{minipage}{0.25\textwidth}
		\myTikzPicture{5.5}{5.5}{black}{red}{black}
	\end{minipage}
	
	\begin{minipage}{0.25\textwidth}
		\myTikzPicture{8}{8}{black}{black}{red}
	\end{minipage}%
	\begin{minipage}{0.25\textwidth}
		\myTikzPicture{7}{7}{black}{red}{red}
	\end{minipage}%
	\begin{minipage}{0.25\textwidth}
		\myTikzPicture{0}{0}{red}{black}{red}
	\end{minipage}%
	\begin{minipage}{0.25\textwidth}
		\myTikzPicture{0}{0}{red}{red}{red}
	\end{minipage}
	
	\caption{
		Combinatorial ground of Theorem \ref{th_donsker}: an illustration of how the shuttering principle of VC theory acts at the core of the arguments encountered in its proof.}
	\label{vc}
\end{figure}
We depict in Figure \ref{vc} the main step in the proof of Theorem \ref{th_donsker}, which essentially involves demonstrating that the class $\left\{1_{\{l \leq t<u\}}(t-l): t \in[0,1]\right\}$ has a VC index $\leq 3$. To show it using a shattering principle, one needs to demonstrate that no set of 3 points can be shattered by the class. This involves proving that for any set of 3 points, there does not exist a subset of functions in the class that can realize all possible dichotomies (i.e., all possible ways to separate the points into two groups). In other words, there must be at least one dichotomy that cannot be represented by any function in the class for any set of 3 points. If this condition holds for every set of 3 points, then the VC index of the class is indeed 3.

\subsubsection{Unbiasedness}
In scenarios involving just right-censoring, it is typically assumed that the censoring time is independent of the survival time, either marginally or conditionally when accounting for external covariates. However, this assumption does not readily extend to cases of interval censoring. In the context of the previously defined notation, where \( L \) and \( U \) denote the interval endpoints and \( T \) represents the survival time, there exists an inherent relationship expressed as \( L < T \leq U \). For interval-censored data, the appropriate assumption to adopt is the \textit{noninformative interval censoring assumption} specified as follows.
\begin{ass}\label{assinfo}	$\mathbb{P}(T \leq t \mid L, U, L<T \leq U, X)=\mathbb{P}(T \leq t \mid L<T \leq U, X).$
\end{ass}
Assumption \ref{assinfo} essentially states that, except for the fact that $T$ lies between $l$ and $u$, which are the realizations of $L$ and $U$, the interval $(L, U)$ (or equivalently its endpoints $L$ and $U$) does not provide any extra information for $T$. In other words, after conditioning on $L$ and $U$ the probabilistic behavior of $T$ remains the same except that the original sample space $T \geq 0$ is now reduced to $l=L<T \leq U=u$. The idea encoded by Assumption 1 is reminiscent of the so-called \textit{conditionally independent censoring assumption} from the right censoring literature \citep{zhang2010interval}.


\begin{propo}\label{prop_unb}
	
	Let Assumption \ref{assinfo} hold. If $F$ is continuous on $t$ for all $x$, then conditional on $X=x, L=l,U=u, L<T \leq U$ 
	$${F}(T,x) \sim \operatorname{Uniform}({F}(l,x),{F}(u,x))$$
\end{propo}

Notice that if $\Phi, \Lambda, \Upsilon$ were built oracle-wise; i.e. using the true conditional distribution function $F$ instead of $\hat F_1$, then if Assumption \ref{assinfo} is true and $F$ is continuous on $t$ for all $x$ we have that $\Phi$ is uniformly distributed on $[\Lambda,\Upsilon]$ conditional on $\Lambda,\Upsilon$ as an inmediate consequence of Proposition \ref{prop_unb}.

For the sake of formal simplicity in the proofs, we will assume that the \(\Phi_i\)'s are genuine probability integral transforms constructed using \(F\), as justified by Proposition \ref{propj} and empirically supported by the evidence in the simulations section. \begin{propo}\label{propj}
	
	Let Assumption \ref{assinfo} hold and let $F$ be continuous on its first argument for all $x$. Let $\operatorname{Uniform}(\hat {F}_1(l,x),\hat {F}_1(u,x))(\cdot)$ be the cdf of the uniform distribution over the interval $(\hat {F}_1(l,x),\hat {F}_1(u,x))$ and define $J(\cdot,l,u,x):={P}\left(\hat{F}_1(T, X) \leq \cdot \mid l \leq T \leq u, X=x\right)$. Then,
	$$ J(\cdot,l,u,x) - \operatorname{Uniform}(\hat {F}_1(l,x),\hat {F}_1(u,x))(\cdot) = \mathcal{O}_\mathbf{P}(r_{N-n})$$
	
\end{propo}

\noindent Note that the $\mathbf{P}$ involved in the stochastic order symbol $\mathcal{O}_{\mathbf{P}}$ represents the marginal probability distribution of split 1. Proposition \ref{propj} asserts that estimating \(\hat{F}_1\) with a non-parametric procedure using split 1 performs nearly as well as constructing oracle scores with the true population conditional survival function.

 The next Corollary ensures that the interval process is unbiased under non-informative interval censoring (Assumption \ref{assinfo}). A proof can be found in the supplement. 
\begin{corollary}\label{a2}
	Suppose Assumption \ref{assinfo} is true and that $F$ is continuous on $t$ for all $x$. Then $E [\mathbb{I}_n(t)]=G(t)$ for any $t \in [0,1]$, where the expectation is taken jointly wrt to $\Phi,\Lambda,\Upsilon$. 
\end{corollary}

Now, in virtue of Theorem \ref{th_donsker}, ensuring the universal-Donsker property for $\mathcal{G}$; and Corollary \ref{a2}, which ensures that the interval process targets the true distribution, we have weak convergence of the centered process.

\begin{corollary}\label{convi}
	Under Assumption 1, the sequence of interval processes $\sqrt{n}\left(\mathbb{I}_n-G\right)$ converges in distribution in the space $D[0, 1]$ to a tight random element $\mathbb{H}_P$, whose marginal distributions are zero-mean normal. 
\end{corollary}

The limit process that shows up in Corollary \ref{convi}, $\mathbb{H}_P$, is a
Brownian bridge because of multivariate central limit theorem: given any finite set of measurable functions $g_i$ with $Pg_i^2<\infty$ then
$\left(\mathbb{H}_n g_1, \ldots, \mathbb{H}_n g_k\right) \rightsquigarrow\left(\mathbb{H}_P(1), \ldots, \mathbb{H}_P(k) \right)
$, where the vector on the right possesses a multivariate-normal distribution with mean zero, provided that Assumption \ref{assinfo} is true and that $F$ is continuous on $t$ for all $x$. 

\subsubsection{Bootstrapping interval processes}
Sampling from $\mathbb{I}_n$ generates a sample $\Phi^*_1, \ldots, \Phi^*_n$ giving rise to the \textit{bootstrap empirical measure}, which corresponds to $ {P}_n^*=n^{-1} \sum_{i=1}^n \delta_{ \Phi_i^*=\Lambda^*_i=\Upsilon^*_i}(\phi, \lambda, \upsilon) $

\noindent and to the \textit{bootstrap empirical process} $\mathbb{H}_n^*=\sqrt{n}\left({P}_n^*-{P}_n\right) \in \ell^{\infty}(\mathcal{G}) $

\begin{remark} Let $t \in [0,1]$. In virtue of Remark \ref{equal} we have ${P}^*_ng_t = n^{-1}\sum_{i=1}^n  1{\{\Phi^*_i \leq t\}} := \mathbb{I}^*_n(t)$.
\end{remark}

\noindent The next result, Corollary \ref{boo}, whose proof is inmediate by using Theorem 23.7 in \cite{vaartas}, is the key formal step of our theoretical arguments, which can be graphically regarded as

\begin{center}
	\begin{tikzcd}[row sep=1.5cm,column sep=2cm]
		\left(\mathbb{I}_n-G\right) \arrow[r, "\mathcal{O}\left(\frac{1}{\sqrt n}\right)"] \arrow[dash,dashed]{d} & \mathbb{H}_P \\
		\left(\mathbb{I}_n^*-\mathbb{I}_n\right) \arrow[ur, "{\mathcal{O}\left(\frac{1}{\sqrt n}\right) \textrm{ cond. } \Phi_1,\Lambda_1,\Upsilon_1, \Phi_2, \Lambda_2,\Upsilon_2 \ldots }"']
	\end{tikzcd}
\end{center}

\noindent or informally writing,

$$\boxed{\sqrt n \left(\mathbb{I}_n^*-\mathbb{I}_n\right) \mid \mathbb{I}_n \stackrel{d}{\approx} \sqrt n \left(\mathbb{I}_n-G\right)}$$

\begin{corollary}\label{boo}
	Conditionally given $\Phi_1,\Lambda_1,\Upsilon_1, \Phi_2, \Lambda_2,\Upsilon_2 \ldots$  the sequence $\sqrt{n}\left(\mathbb{I}_n^*-\mathbb{I}_n\right)$ converges to $\mathbb{H}_P$. 
\end{corollary}

We finally present the two main results regarding consistency of \texttt{uncervals}, which form a crucial part of this paper.  {They essentially guarantee that the prediction set  $\widehat{ \mathcal{C}^e}_{1-\alpha} (X_{N+1}) $, which is the output of Algorithm 1, will asymptotically exclude the true value of the interval-censored target for $e=*$ with a probability that is at least as large as the user-specified confidence level. }

\begin{theorem}[Asymptotic unconditional validity]\label{uncond} Let Assumption 1  hold. Then,
	$$
	\mathbb{P}\left(T_{N+1} \in \widehat{\mathcal{C}}_{1-\alpha}^{* }\left(X_{N+1}\right)\right)=1-\alpha+o_\mathbb{P}(1) .
	$$
\end{theorem}

\begin{theorem}[Asymptotic conditional validity] \label{asycond} Let Assumption 1 and hold. Suppose that $F(T,X) \perp  X$ (i.e., no misspecification). Then,
	$$
	\mathbb{P}\left(T_{N+1} \in \widehat{\mathcal{C}}_{1-\alpha}^{*}\left(X_{N+1}\right)\mid X_{N+1}\right)=1-\alpha+o_\mathbb{P}(1) .
	$$
	
\end{theorem}

\begin{remark} We note that the probability $\mathbb{P}$ involved in Theorems \ref{uncond} and \ref{asycond} (unconditional and conditional on $X_{N+1}$ respectively) is the one from which the entire dataset $\mathcal{D}_N$ and the new data point $(X_{N+1},Y_{N+1})$ were originally sampled. 
	
\end{remark}

The proof of Theorem \ref{asycond} is the same as that of Theorem \ref{uncond} having into account that under correct specification, ranks are independent of the predictors unlike regression residuals. Theorems \ref{uncond} and \ref{asycond} establish the asymptotic validity of our procedure under weak and easy to verify conditions. Also, Theorem \ref{asycond} implies that under its assumptions, Algorithm \ref{oa} outputs an interval that asymptotically coincides with the oracle when $b=\frac{1}{2}$. 

\subsubsection{Convergence rates}

 {In this section, we express the convergence rate of the coverage to the desired level as a function of the rate at which $\hat F_1$ is estimated. Let $K$ be the cumulative distribution function of the oracle scores $|F(T_i,X_i) - b|$, $\hat K(v) = \frac{1}{n}\sum_{i=1}^n 1\{|\hat F_1(T_i,X_i) - b|<v\}$ and $\tilde K(v) = \frac{1}{n}\sum_{i=1}^n 1\{|F(T_i,X_i) - b|<v\}$. }

\begin{ass}\label{survrate} There exist $0<\tau<\infty$ and $r_{N-n}$ such that
	$\sup _{t \leq \tau, x \in \mathbb{R}^p} \left|\hat{F}_1\left( t,x\right)-F\left(t,x\right)\right| = \mathcal{O}_\mathbf{P}\left(r_{N-n}\right)
	$
	
\end{ass}

\noindent We note that $N-n$ is the sample size of split 1, the data used to estimate $\hat F_1$. For example, under certain assumptions,  \cite{cho2023multi} derive the following uniform rate of convergence for single-draw-per-person random survival forest (technical details regarding $n_{min}, \alpha, \varphi$ and $d_l$ can be found in that article)
$$r_{N-n}^{\prime}=\max \left\{\sqrt{\frac{\log (N-n)\left\{\log \left(n_{\min }\right)+\log \log (N-n)\right\}}{n_{\min }}},\left(\frac{n_{\min }}{N-n}\right)^{\frac{\log ((1-\alpha)-1)}{\log \left(\alpha^{-1}\right)} \frac{0.991 \varphi}{\max _l d_l}}\right\}$$

The following result is derived from an adaptation of the proof of Lemma 1 in \cite{cherno}. 

\begin{lemma}\label{rate} Assume that $K$ is such that $\sup _{v_1 \neq v_2}\left|K\left(v_1\right)-K\left(v_2\right)\right| /\left|v_1-v_2\right| :=W < \infty$, i.e. $K$ is $W$-Lipschitz. Then we have for any $\delta > 0$
	$$\begin{aligned}\sup _{v \in \mathbb{R}}\left|\widehat{K}(v)-K(v)\right|\leq 2 \frac{\sup _{t \leq \tau, x \in \mathbb{R}^p}\left|\hat{F}_1(t, x)-F(t, x)\right|^2}{\delta^2}+2 \delta W+3 \sup _{v \in \mathbb{R}}|\widetilde{K}(v)-K(v)| \quad \mathbb{P}-a.s.
	\end{aligned}$$
\end{lemma}


 Next, we use the classical Glivenko-Cantelli Theorem for the part involving \(\widetilde{K}\) in Lemma \ref{rate}, and for the rest, we choose \(\delta\) optimally to tighten the bound in Lemma \ref{rate} as much as possible by basic univariate differentiation.

\begin{propo}\label{prepropo}Let Assumption \ref{survrate} hold and let $K$ be $W$- Lipschitz. Then, 
	$$\sup _{v \in \mathbb{R}}\left|\widehat{K}(v)-K(v) \right| = \mathcal{O}_{\mathbf{P}}\left((Wr_{N-n})^{2/3}\right)  + \mathcal{O}_{P}\left(n^{-1/2}\right) $$

\end{propo}

\begin{corollary}[Convergence rate of \texttt{uncervals}] \label{finalconvrate} Let Assumptions 1,2 and 3 hold. Also, let $K$ be Lipschitz. Then,
	$$
	\mathbb{P}\left(T_{N+1} \in \widehat{\mathcal{C}}_{1-\alpha}^{*}\left(X_{N+1}\right)\mid X_{N+1}\right)=1-\alpha+O_\mathbb{P}\left(\operatorname{max}\{r^{2/3}_{N-n},n^{-1/2}\}\right) .
	$$
	
\end{corollary}

 {This rate is reminiscent of the recent insights from bagging \citep{bagging}, where averaging over perturbations of the data introduces stability and robustness, even if it does not always improve asymptotic rates.Moreover, Corollary \ref{finalconvrate} is derived using the bound in Proposition \ref{prepropo}, which relies on the supremum norm and results in a lack of tightness in its subsequent applications. Thus, the rate captures the practical benefits of conformal inference, which are critical for finite-sample performance.  }

\subsection{Technical considerations}
Consider a covariate-free setup where only right censoring takes place, for simplicity. Consider survival functions $S$ with $S(0)=1$ and restricted to $[0, \tau]$, where $\tau<\infty$  is such that neither the survival function of the true times nor the one of the censoring times is zero coming from the left. Then $\hat S$ is uniformly consistent for $S$ over $[0,\tau]$, where $\hat S$ is the Kaplan-Meier estimator defined as $\hat{S}(t)=\prod_{i:T_i \leq t}\left(1-\frac{d_i}{n_i}\right)$, where $T_i$ denote observed events, $d_i$ denote number of deaths that happen at $T_i$ and $n_i$ denote number of individuals at risk just before $T_i$ (units that have been censored are not considered to be at risk), see \cite{kaplan1958nonparametric,kosorok}.  The Kaplan-Meier estimator does not reach zero beyond the last observed event time if there are still individuals at risk who have not experienced the event (right-censored observation).  Beyond the last event time, there are no further decreases in the estimator, leading to a tail that does not drop to zero. Using predictive regions of the form $\left\{t: b-\hat{Q}_{\mathcal{I}_2}^* \leq \hat{F}_1(t) \leq b+\hat{Q}_{\mathcal{I}_2}^*\right\}$ can result in a confidence interval with an infinite upper bound lead to LPBs because there might not exist $0<t<\tau$ such that $b+ \hat{Q}_{\mathcal{I}_2}^* < \hat{F}_1(t)$. This inconsistency arises when the desired regions' image under $\hat{F}_1$ is centered around $b <1$ (for example, around the median when $b=\frac{1}{2}$). Recall that setting $b=1$ when using \texttt{uncervals} produces predictive regions for survival time of the form $\left\{y: \hat{S}_1(y) \leq \hat{Q}_{\mathcal{I}_2}^*\right\}$, ensuring that the definition of the confidence region is not contradicted despite the tail finite-sample behaviours of the Kaplan-Meier estimator.

\section{Simulation studies} \label{sec:simulation}

In this section, we conduct simulations to empirically evaluate \texttt{uncervals}. First, we study its behaviour in terms of valid conditional coverage and empirically show that naively considering $\hat F^{-1}_1 (\alpha, \cdot )$ as an $(1-\alpha)$ LPB is a completely miscalibrated choice. In Section \ref{sim:asym} we empirically verify that the coverage probability of the prediction intervals with $b=1/2$ and $e=\texttt{*}$ converges to the nominal level \(1 - \alpha\) as the sample size increases, as predicted by theoretical results. Next, in Section \ref{sim:lpbs} we compare three approaches for constructing lower prediction bounds (LPBs): a naive quantile-based method, \texttt{uncervals} using $b=1$ and $e=\texttt{0}$, and \texttt{uncervals} operating with $b=1$ and $e=\texttt{*}$, highlighting their performance differences and trade-offs. Finally, we assess the Conformalized Survival Analysis method by \cite{candesrssb} in Section \ref{evalcfs}, demonstrating its high sensitivity to the choice of the threshold parameter \(c_0\). In the supplement, we mathematically formalize how the \texttt{simIC\_weib} function from the \texttt{icenReg} package, tailored for regression with interval-censored targets, operates under an accelerated failure time (AFT) model with a Weibull distribution. This provides the basis for simulating interval-censored outcomes and their covariates across various scenarios.

 {We emphasize that in three out of four sections of the simulations presented in this article, the data is interval-censored. Right-censoring appears only in Section \ref{evalcfs}, as the benchmark \citep{candesrssb} is limited to operate under that type of censoring.}

\subsection{Conditional coverage}\label{sim:cond}

We simulate $B=100$ datatsets with \texttt{par\_shape=2, par\_scale=1, par\_inspections=10},  $n=500$, $X \sim \operatorname{Uniform}(-2,2)$ and $r(X)=-0.3 |X|$ (see the Supplement for more details on what these values represent in the simulation of interval-censored data). In all cases, we chose the base algorithm to be Interval Censored Recursive Forests. For each of these runs, we obtain two LPBs: one given by \texttt{uncervals} with $e=*$ and the other by the naive quantile approach. We simulate $5000$ fresh datapoints following the same mechanism and we check if the true survival time was higher than the LPB. For each approach, this leads to a dataset $(1\{T_1 \geq \widehat L_{1-\alpha} (X_{1})\},X_1), \ldots, (1\{T_{5000} \geq \widehat L_{1-\alpha} (X_{5000})\},X_{5000})$. We fit a logistic regression using \texttt{gam} from the R library \texttt{mgcv} \citep{mgcv} so that we obtain the curves visible in Figure \ref{fig:condcov}, which are estimations of the success probability $ \pi (X_{N+1}) = \mathbb{P}(T_{N+1} \in \hat C_{1-\alpha}(X_{N+1}) \mid X_{N+1})$ provided by $\texttt{gam}$. We consider the empirical version of the square root $L_2$-error $\sqrt{\int_{-2} ^2( \widehat \pi (x) - (1-\alpha)) ^2 dx}$ given by $\texttt{err}=\sqrt{\frac{1}{5000}\sum_{i=1}^{5000} (\hat \pi (X_i) - (1-\alpha)) ^2}$. 

\begin{figure}[t!]
    \centering
       \begin{minipage}{0.6\textwidth}
        \centering
\resizebox{0.9\textwidth}{!}{ 
      \begin{tabular}{lcc|cc|c}
            \hline
            & \multicolumn{2}{c|}{$\alpha=0.1$} & \multicolumn{2}{c|}{$\alpha=0.05$} & \\
            \hline
            & Mean & SD & Mean & SD & \\
            \hline
            Naive & 8.30 & 2.29 & 7.30 & 1.98 &  \multirow{2}{*}{$n=500$} \\
            \texttt{uncervals} & 6.52 & 2.26 & 4.99 & 1.54 & \\
            \hline
            Naive & 7.49 & 1.43 & 6.79 & 1.76 &  \multirow{2}{*}{$n=1000$} \\
            \texttt{uncervals} & 4.97 & 1.51 & 4.25 & 1.35 & \\
            \hline
        \end{tabular}
       }
        \caption{Summary statistics for \texttt{err} $\cdot 100$. }
        \label{tab:summary}
    \end{minipage}%
     \begin{minipage}{0.4\textwidth}
        \centering
        \includegraphics[width=\linewidth]{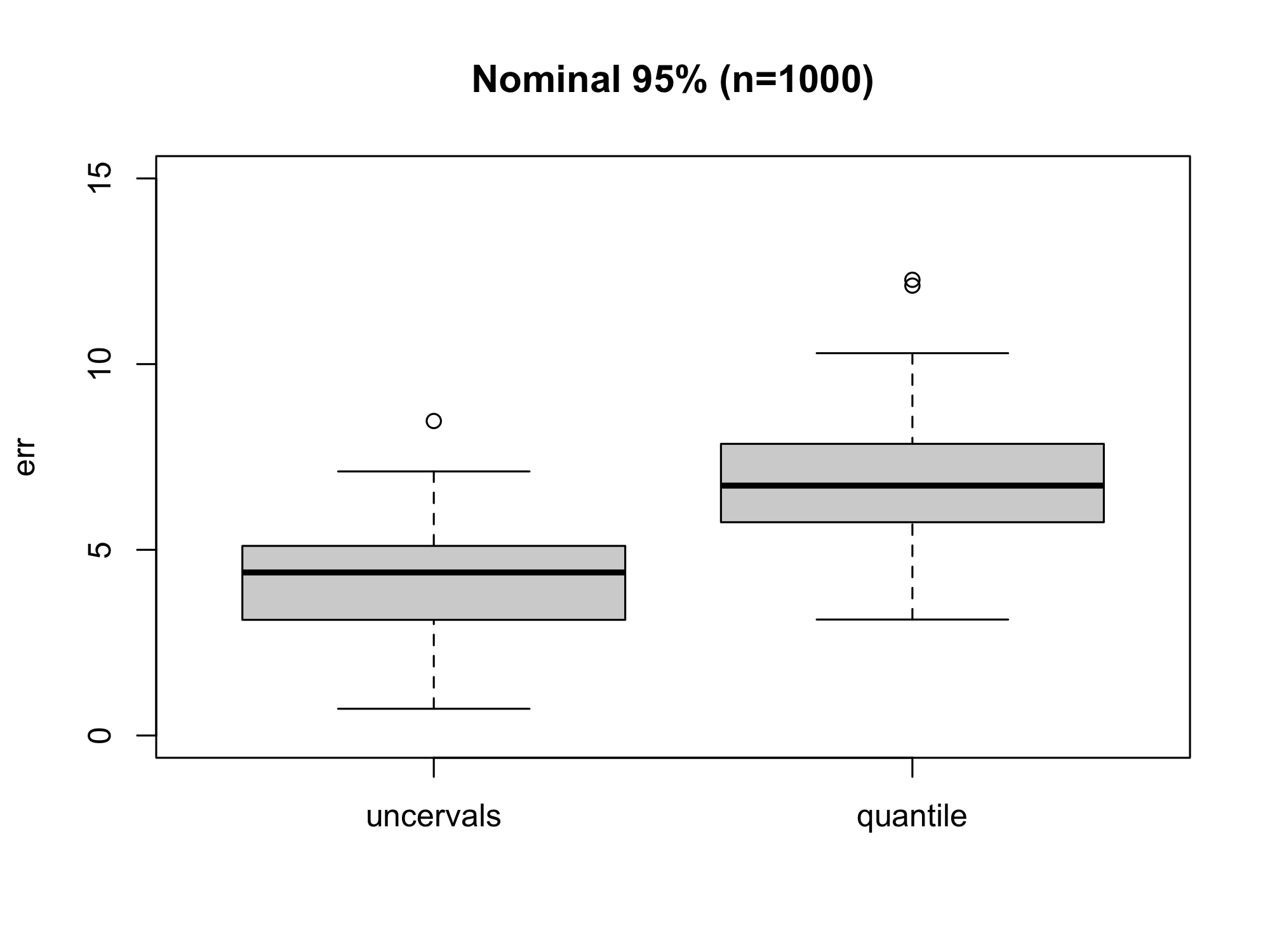}
        \caption{Boxplots for the distribution of \texttt{err} $\cdot 100$ across simulation runs. Additional cases in the Supplement. }\label{fig:boxp}
    \end{minipage}
\end{figure}

Average and standard deviations of \texttt{err} across $B$ simulation runs are displayed in Figure \ref{tab:summary}. We observe that our approach reduces the average \texttt{err} by $30\%$ for $\alpha = 0.1, n = 500$; $50\%$ for $\alpha = 0.05, n = 500$ and  $\alpha = 0.1, n = 1000$ ; and $60\%$ for $\alpha = 0.05, n = 1000$. Note that when setting $\alpha = 0.95$, we construct predictive regions intended to enclose $95\%$ of the data. In this context, a $60\%$ reduction in conditional coverage error is highly significant. Looking at the standard deviations, ours is always lower, which is an indicator of improved stability. This is also observed in Figure \ref{fig:condcov}, which shows that, apart from being miscalibrated, the quantile approach is significantly more unstable across the support of the covariates.

\begin{figure}[h!]
	\centering
		{\includegraphics[width=0.45\linewidth]{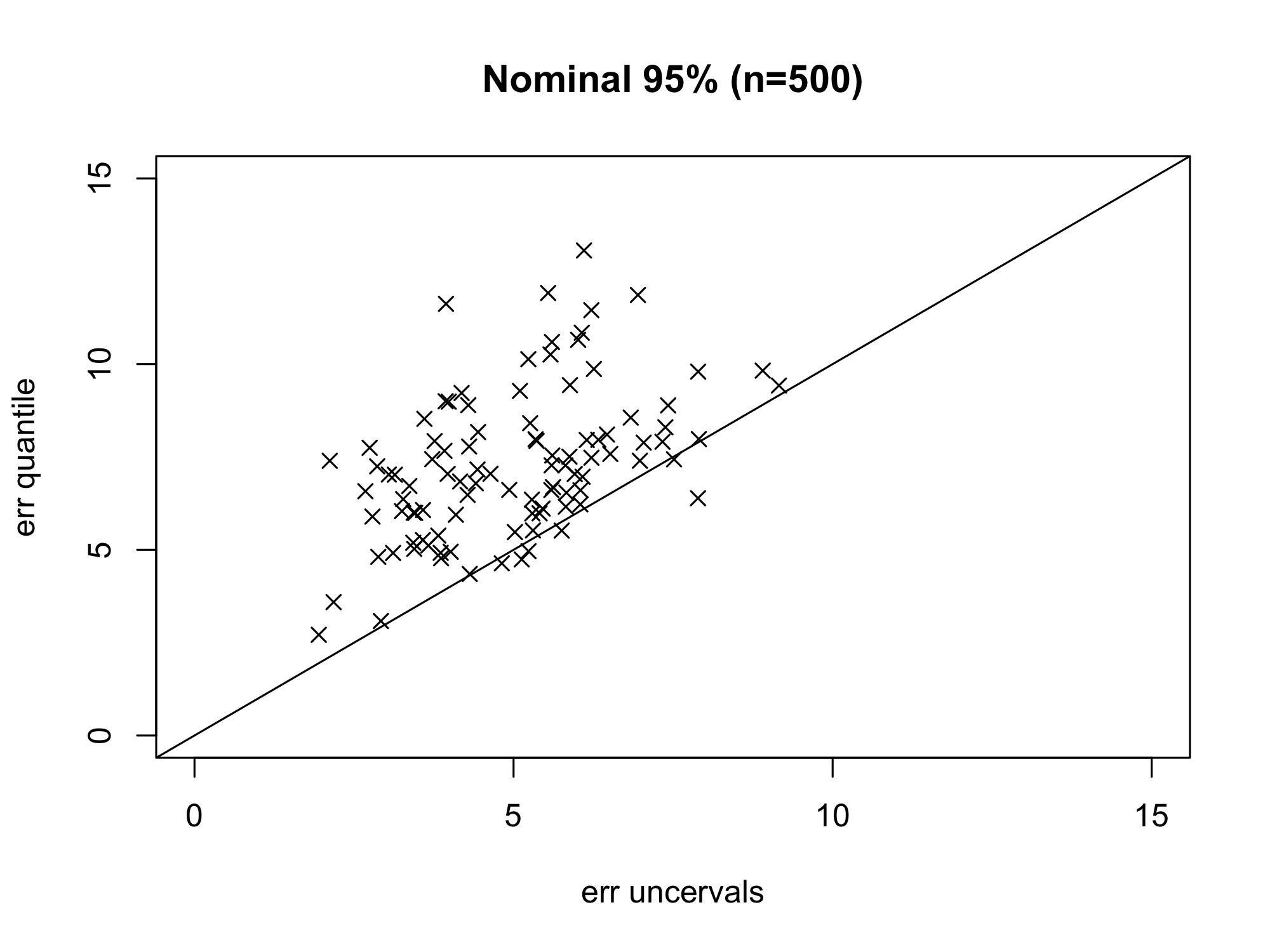}}\quad
	{\includegraphics[width=0.45\linewidth]{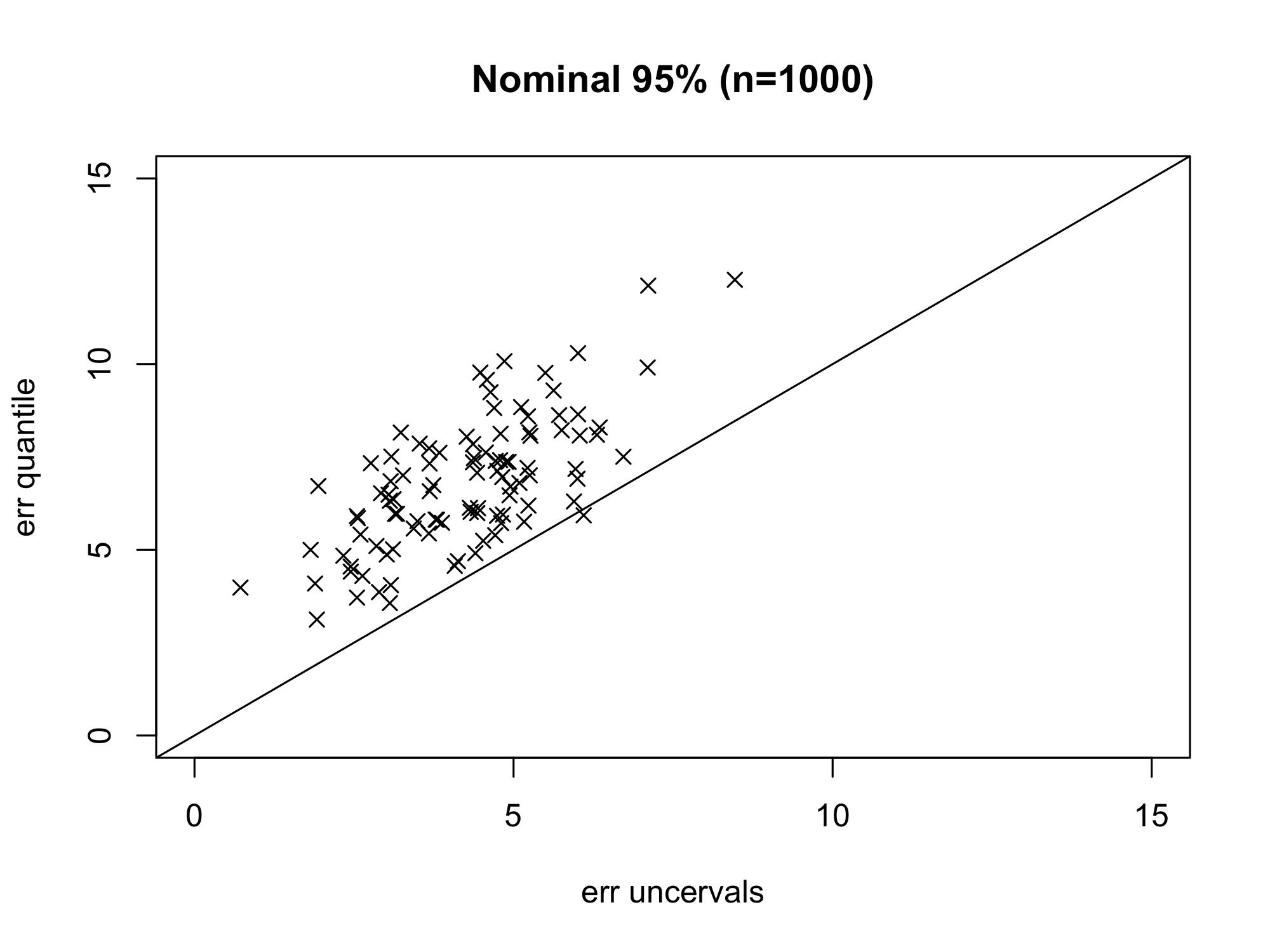}}
	\caption{ Becnhmark plots for our approach vs naively taking the $\alpha$- th quantile considerin \texttt{err} as a metric. Notice how $\texttt{uncervals}$ performs better most of the times, especially when increasing the sample size to $n=1000$. See plots for $\alpha=0.1$ in the Supplement }
	\label{fig:best}
\end{figure}

In both plots in Figure \ref{fig:best}, the x-axis represents \texttt{err} for our approach, while the y-axis represents \texttt{err} for the naive quantile method. Each cross corresponds to a simulation run. A cross above the diagonal indicates that, in that particular run, the naive quantile method had a greater \texttt{err} than \texttt{uncervals}. Since most points lie above the diagonal, this suggests that our approach outperforms the naive method with high probability. This is also evident from the discrepancy between the two boxplots in Figure \ref{fig:boxp}.


\subsection{Verfication of asympotic properties}\label{sim:asym}
\label{sim_asym}
We empirically corroborate what Theorem \ref{uncond} implies: the probability that the new observation $T_{N+1}$ falls within the prediction interval $\widehat{\mathcal{C}}_{1-\alpha}^{*}(X_{N+1})$ is approximately $1 - \alpha$ as the sample size grows. In this setup, \texttt{uncervals} is working under $e=*$ operating mode, i.e. we uniformly randomize $\Phi^*$ over the interval whose upper and lower borders are $\hat F_1$ evaluations on the original intervals. 

The results are visible in the Supplement. When we have more data, variability decreases in general.  The more extreme $\alpha$ is, the slower convergence is. As $1-\alpha$ decreases, this effect becomes more pronounced. It should also be noted that estimating \(\hat{F}_1\) using Interval Censored Recursive Forests (setup c), as would typically be done in real applications, performs almost as well as using the true population conditional survival function (setup d).

\subsection{Comparison of different LPBs}\label{sim:lpbs}

We consider three different sample sizes, numbers of covariates and confidence levels. We propose  $\alpha_1=0.8$, $\alpha_2=0.9$, $\alpha_3=0.95$. In all cases, we set $b=1$ (therefore LPB). For each situation, we evaluate the following three different algorithms for uncertainty quantification: 1. checking for which $t>0 $ whether $\hat{F}_1(t, X_{\text{N+1}})$ is greater than $\alpha$: the LPB is $\hat F_1^{-1}(\alpha)$ (``naive quantile''); 2. our approach with $e=0$, this is, considering evaluations of $\hat{F}_1$ just on the left borders, thus a trivial extension of \cite{candesrssb} to interval censoring. Instead of randomizing $\Phi_i^* \leftarrow \texttt{runif}\left(\hat{F}_1\left(L_j, X_j\right), \hat{F}_1\left(U_j, X_j\right)\right)$ we set $\Phi_i^* \leftarrow \hat{F}_1\left(L_j, X_j\right)$ so that the finite sample coverage bound holds; and 3. our approach with $e=*$ (randomizing between border evaluations).

\noindent The results of miscoverage across different sample sizes, number of dimensions and significance levels are visible in the Supplement, reescaled by 100. For moderate sample sizes, since we are neither in the small sample size regime nor in the asymptotic regime, \texttt{uncervals} prevails. The performance of our approach is also superior when the dimensionality increases.

\subsection{Evaluating Conformalized Survival Analysis}\label{evalcfs}

In this section, we perform a sensitivity analysis of the \texttt{cfsurv} function in the \texttt{cfsurvival} package from the Conformalized Survival Analysis paper \citep{candesrssb}. There, a threshold $c_0$ is introduced so that the focus is put on subpopulations where $C \geq c_0$. A distributional shift between subpopulations and the whole population arises because patients with larger censoring times tend to be healthier, leading to different joint and conditional distributions of the variables $X$, $C$, and $T$. The authors suggest using a secondary censored outcome $T \wedge c_0$ and highlight that while there is a covariate shift, this can be adjusted by reweighting the samples through weighted conformal inference \citep{weighted}, allowing for a calibrated lower prediction bound (LPB) on $T \wedge c_0$ and thus on $T$. This approach reduces the power loss due to censoring, but depends on choosing the threshold hyperparameter $c_0$.  Performance in terms of empirical coverage depends strongly on the chosen $c_0$, see in Figure \ref{fig:c00}  how empirical coverage degrades with the hyperparameter $c_0$. We simulate as in Section \ref{sec:simulation} but now setting \texttt{par\_shape=2, par\_scale=1, par\_inspections=5},  $n=2000$, $X \sim \operatorname{Uniform}(0,2)$ and $r(X)=X$ without left censoring (see supplement); with $\alpha=0.1$. Under this configuration, right-censoring amounts to approximately $30 \%$. For a grid of $c_0$ ranging from $0.1$ to $1$ we launch conformalized survival analysis by setting $\tilde{T}$ to the lower bounds, the event indicator to $1 \{U=+\infty\}$ and the censoring times to $1 \{U=+\infty\}L + 1 \{U<+\infty\} U$.  Also, recall that the methodology in \cite{candesrssb} requires observing $C$ even when survival time $T$ is available.

\section{Empirical evaluation and comparison}\label{sec:realresults}

\subsection{Systolic Blood Pressure Intervention Trial (SPRINT)}

In order to assess the performance of our methodology, we use data from the NIH's Systolic Blood Pressure Intervention Trial (SPRINT) \cite{sprint}. The primary goal of this analysis is to relate sleep time to cardiovascular diseases. In this example, we synthetically generate a censoring time \(C > 0\) (which is just needed by \cite{candesrssb} whereas not by our approach) that has been simulated according to \texttt{rexp(ninitial, rate = 0.01)}, together with natural data as explained in Table \ref{table:wakeup}. 



In our analysis, we examine the relationship between ``Time that sleep ended'' and ``Morning Systolic BP''. This relationship is pertinent because morning BP can influence wake-up times due to the physiological stress or discomfort associated with elevated BP levels. Understanding how morning systolic BP affects when individuals wake up can provide valuable insights into the interplay between BP management and sleep quality.

We define two clinical outcomes: an interval-censored outcome to use interval-censored algorithms, and a right-censored outcome to use the existing research on conformal prediction \citep{candesrssb}. We assume that participants misunderstood their instructions and did not record the exact time they woke up. However, we have access to the time they set their alarms for, and they remember whether they woke up naturally or with the alarm. We craft the interval-censored $(L, U)$ and the right-censored $(\widetilde{T}, \text{Event indicator})$ data following Table \ref{table:wakeup}.

\begin{table}[h!]
	\centering
	\begin{tabular}{|l|l|c|c|c|c|c|c|}
		\hline
		& &\textbf{L} & \textbf{U} &$\mathbf{\tilde{T}}$ & \textbf{Event indicator} & \textbf{C} & \textbf{T} \\
		\hline
		Natural wake up & \texttt{(zzzzz)--!}& \texttt{(}& \texttt{!} &\texttt{(} & 1 & \texttt{!} &  \texttt{)}\\
		\hline
		Alarm wake up & \texttt{(zz!--)}&\texttt{!}  &$+\infty$ & \texttt{!}&0& \texttt{!} &  \texttt{)} \\
		\hline
	\end{tabular}
	\caption{Casuistic that has been followed in order to build the semi-synthetic data. The original interval-censored observations were given by two numbers, \texttt{'('} and \texttt{')'} , representing the time when sleep started and ended respectively. \texttt{!} encodes an artificial independent censoring time, ensuring that the new target amounts to 10\% right-censoring approximately.}
	\label{table:wakeup}
\end{table}

The final dataset analyzed contains a total of 863 individuals, which we randomly split into 80\% training and 20\% testing sets. First, we consider the approach in \cite{candesrssb}, using a \cite{cox72} model to fit the conditional survival function. We feed it with semi-synthetic training data, which are samples from $(X, \widetilde{T}, \text{Event indicator}, C)$, where $X$ represents \texttt{MORNING\_SYSTOLIC}. Since $\widetilde{T} \leq T$, any calibrated lower predictive bound on the censored survival time $\widetilde{T}$ is also a calibrated lower predictive bound on the uncensored survival time $T$. We try several values of $c_0$, but the coverage is always 1, indicating overly conservative behavior.

Next, we make \texttt{uncervals} operate under \texttt{e=*}. For this, we use \texttt{ic\_sp(Surv(L, U, type = 'interval2'), model = 'ph')} from \texttt{icenReg} package to fit $\hat{F}_1$. We repeat the procedure $B = 100$ times by randomizing the train-test split and, within the training set, the split used in the two phases of conformal inference. We obtain a median coverage of 0.8576, indicating that the Cox model may be misspecified in this setting. Next, we try Interval Censored Recursive Forests another $B = 100$ times and observe a median coverage equal to 0.93.

In conclusion, this example illustrates with real data that the application of Conformalized Survival Analysis can lead to unpractical conservative solutions in practice and that we must use specific data analysis tools that exploit the interval nature of the outcome as \texttt{uncervals} do. 


\subsection{NHANES Physical Activity Example}

The goal of this analysis is to demonstrate, through a clinical application of time-to-event analysis for interval-structured outcomes, how quantifying uncertainty in the response can provide new clinical insights. In this case, our findings are drawn from the fields of public health and physical activity literature. We focus on analyzing data from the NHANES U.S. population study.

The National Health and Nutrition Examination Survey (NHANES) is a comprehensive program conducted by the National Center for Health Statistics (NCHS), primarily aimed at collecting health and nutrition data from the United States population. In this study, we utilized a subset of \(n=2977\) patients aged between 50 and 80 years, gathered from NHANES waves 2003-2006. This subset contains high-resolution physical activity information measured by accelerometer devices over 3-7 days.

Our off-the-shelf semiparametric model in this case is the following vanilla \cite{cox72} model: $h(T \mid \text{Age}, \text{TLAC}) = h_0(T) \exp(-\text{Age}\beta_{\text{Age}} - \text{TLAC}\beta_{\text{TLAC}})$

where \( h(\cdot) \) denotes the baseline hazard function, \( T \) is patient survival in months, \(\text{Age}\) is measured in years, and \(\text{TLAC}\) is a variable related to physical activity levels (with higher physical activity corresponding to higher \(\text{TLAC}\) values). We use \texttt{uncervals} with \(e=*\) and a confidence level of \(\alpha=0.95\). Table \ref{tabla:uncert} shows the $95\%$ LPBs for different combinations of Age and TLAC variables corresponding to the 20th and 80th percentiles of these covariates. 

\begin{table}[ht!]
	\centering
	
	\begin{tabular}{lcc}
		\toprule
		& \textbf{TLAC} $=2263$ & \textbf{TLAC} $=3241$\\
		\textbf{Age} $=58$ & $(87.89, +\infty)$ & $(124.24, +\infty)$ \\
		\textbf{Age} $=73$ & $(45.45 , +\infty)$ & $(62.12,+\infty)$\\
		\bottomrule
	\end{tabular}
    \caption{Intervals whose left boundaries are the estimated LPBs for different values of the covariates Age and TLAC}
	\label{tabla:uncert}
\end{table}

For younger individuals (Age = 58) and more active individuals (TLAC = 3241), the analysis indicates they can live at least 124 months from study onset with high probability. Conversely, older individuals (Age = 73) and less active people (TLAC = 2263) have a threshold of 45.45 months, which is approximately one-third of the upper bound for the previous setting. In intermediate cases, there is a 50\% increase in survival time with higher TLAC values and a 100\% increase when reducing the age to 58 years, approximately. This suggests that age has a stronger influence on the lower predicted survival bound compared to TLAC.

\begin{figure}[t!]
	\centering
	\begin{subfigure}[t]{0.48\textwidth}
		\centering
		\includegraphics[width=\textwidth]{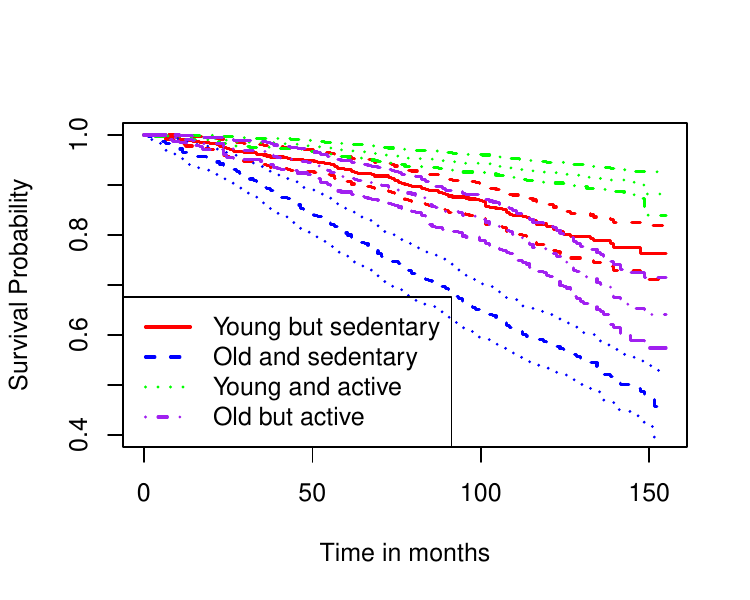}
		\caption{Estimated survival functions for four different groups in the NHANES data. Each marginal Turnbull estimator is fed with data from one of the four subsets, created by partitioning the dataset based on whether individuals' age and TLAC are below or above the medians of these covariates. Note that older individuals who engage in sufficient physical activity exhibit survival probabilities comparable to those of younger individuals who lead sedentary lifestyles.}
		\label{fig:c0}
	\end{subfigure}
	\hfill
	\begin{subfigure}[t]{0.48\textwidth}
		\centering
		\includegraphics[width=\textwidth]{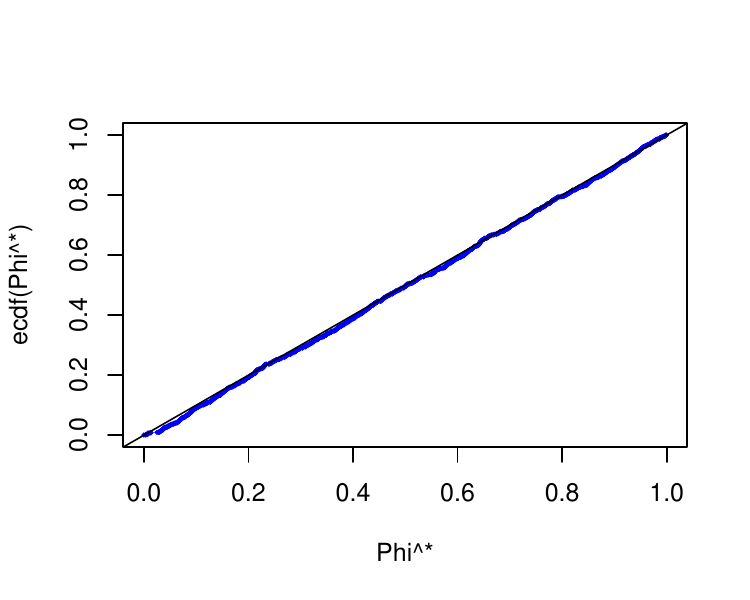}
		\caption{In blue, empirical cumulative distribution function of $\{\Phi^*_i\}_{i \in \mathcal{I}_2}$ using the NHANES database. The \texttt{ic\_sp} algorithm (which uses a Cox model) was utilized as the base procedure to compute $\hat{F}_1$. For comparison, the black line illustrates the theoretical cumulative distribution function of a uniform distribution $\operatorname{Uniform}(0,1)$, which is the identity function.}
		\label{fig:good}
	\end{subfigure}
	\caption{Results for NHANES dataset.}
\end{figure}

Our results in Table \ref{tabla:uncert} and Figure \ref{fig:c0} suggest that the uncertainty of survival depends on patient characteristics so that informed decisions should incorporate public health criteria based on uncertainty rather than relying solely on point estimates, such as expected survival time, to achieve more refined clinical decisions. For example, a 58-year-old inactive individual will have a life expectancy that is 50\% shorter than that of an active individual, with a probability greater than 95\%. In the U.S. population, for instance, only about 5\% of those who are very active at age 58 will die before reaching 21 additional years of life. The current average life expectancy in the U.S. population is approximately 75 years, and these results provide new insights into the survival of patients who are at least 58 years old.

\subsubsection{Goodness-of-fit in for interval-censoring}
The goal of this subsection is to illustrate how we can exploit \texttt{uncervals} to evaluate whether using a Cox model in the NHANES case application is well specified via a goodness-of-fit approach. The general idea of extending such a procedure is similar to \cite{fernandez2019maximum}, but in that research the authors focus solely on right-censored data.

Assuming that the random sample $\mathcal{D}_{N}$ is distributed according to a theoretical conditional cdf $F_0$, under the null hypothesis $H_{0}: F=F_{0}$ we have $F\left(X_{i},T_i\right) \sim \operatorname{Uniform}(0,1)$ in virtue of the probability integral transform theorem provided that $F_0$ is continuous. Therefore, testing the null hypothesis is equivalent to testing $H_{0}: F \circ F_{0}^{-1} = F_{\text{unif}}$, where $F_{\text{unif}}$ denotes the uniform distribution function on $(0,1)$. Notice that since we have interval-censored data, we do not observe the failure times $T_{i}$ but instead observe $(L_i, U_i)$. Under non-informative interval-censoring (Assumption 1) and absence of misspecification (Assumption 2), the empirical distribution function $\frac{1}{n} \sum_{i=1}^{n} 1\{\Phi^*_i \leq \cdot\}$ to that of a $\operatorname{Uniform}(0,1)$ under the null hypothesis- recall that $\Phi^*_i = \hat{F}_1(L_i, X_i) + \textrm{Uniform}(0,1) \cdot (\hat{F}_1(U_i, X_i) - \hat{F}_1(L_i, X_i)).$

A natural application in our NHANES application case arises, which is testing whether the Cox model is well-specified by evaluating the pseudo-random scores $\Phi^*_i \sim \operatorname{Uniform}(0,1),\ i=1,\dots,n$ via a graphical criterion (as shown in Figure \ref{fig:good}). In this case, we can see that the method is well-calibrated and there is no evidence that the conditional survival function class is misspecified with a Cox model. A Kolmogorov-Smirnov test would have to be performed in order to conduct a formal hypothesis test. 


Users might notice by inspection of Figure \ref{fig:c0} that using the constructed $\Phi^*_i$'s based on taking $\hat{F}_1=$ Turnbull's estimator- i.e. ignoring the covariates- seems to be correct and might lead one to believe that it is acceptable to disregard the predictors. This is entirely expected, as Turnbull's estimator is fully non-parametric and, therefore, will naturally produce residuals with the correct distribution. However, what we are testing is whether the parametric specification formulated in the Cox model is correct for the two chosen covariates: Age and TLAC.


\section{Discussion}\label{sec:discuss}

The present paper proposes a new framework for uncertainty quantification in the context of interval-censored data. We introduce novel algorithms along with new empirical process tools to examine the theoretical properties of the methods presented. In certain special cases, the methods introduced here are conformal algorithms that exhibit non-asymptotic guarantees of the form $\mathbb{P}(T_{N+1} \in \widehat{\mathcal{C}}_{1-\alpha}(X_{N+1})) \geq 1 - \alpha$. These proposed methods are a natural progression of those proposed by \cite{candesrssb} for right-censored data, extending them to interval-censored data. In this sense, \texttt{uncervals} can be seen as a trade-off between the work by \cite{candesrssb} and that of  \cite{politis2015model}. Diverse numerical results demonstrate clear advantages of our approach in terms of coverage approximation.

For the theoretical analysis, we introduce a new class of functions that allows to derive new asymptotic properties of the uncertainty quantification methods presented. We illustrate the advantages of our proposal over a naive adaptation for interval-censored data of \cite{candesrssb} in a clinical problem related to sleep disorders. Additionally, we highlight the application and scientific interest of these methods in quantifying the impact of physical activity on patient survival. We also introduce the core steps to use the interval-censored methodology to do goodness of fit in this setting. 

As future work, we suggest extending the proposed framework to handle truncated or double-truncated censored data (see \cite{https://doi.org/10.1002/sim.9828}). Additionally, we propose examining the methods introduced here for multivariate interval data. We aim to develop tolerance regions through resampling techniques by exploiting the interval process and its corresponding universal-Donsker property. To the best of our knowledge, this would be the first general framework for tolerance regions \citep{li2008multivariate} for interval-censored data.


\bibliographystyle{agsm}
\bibliography{Bibliography-MM-MC}

\appendix
\section{Proof of Theorem 2}
The elements in $\mathcal{G}$ are well-defined. For $t \geq u$ the numerator is $u-l$, which yields the ratio $1$. For $t < u$, the ratio is bounded by $1$. 
First note that
$$
g_t(l, u)=1_{\{0 \leq l \leq u \leq 1\}} 1_{\{u \leq t\}}+1_{\{l \leq t<u\}}(t-l) \times 1_{\{0 \leq l<u \leq 1\}}(u-l)^{-1},
$$
and thus $\mathcal{G} \subset \mathcal{F}_1+\mathcal{F}_2 \cdot \tilde{g}(l, u)$, where $\mathcal{F}_1=\left\{1_{\{0 \leq l \leq u \leq 1\}} 1_{\{u \leq t\}}: t \in[0,1]\right\}, \mathcal{F}_2=\left\{1_{\{l \leq t<u\}}(t-l)\right.$ : $t \in[0,1]\}$, and $\tilde{g}(l, u)=1_{\{0 \leq l<u \leq 1\}}(u-l)^{-1}$. Standard arguments yield that $\mathcal{F}_1$ is a Donsker class. We will next show that $\mathcal{F}_2$ is a $\mathrm{VC}$ class with $\mathrm{VC}$ index $\leq 3$. This then implies that $\mathcal{F}_3 \equiv \mathcal{F}_2 \cdot \tilde{g}$ is also $\mathrm{VC}$ with $\mathrm{VC}$ index $\leq 5$ by part (vi) of Lemma 9.9 of \cite{kosorok}. We will then show that $\mathcal{F}_3$ is also pointwise measurable. This, combined with Proposition 8.11 (to establish sufficient measureability), Theorem 9.3 (to establish boundedness of the uniform entropy integral), and Theorem 8.19 of \cite{kosorok}, yield that $\mathcal{F}_3$ is Donsker. Since the sum of two Donsker classes is also Donsker, the desired conclusion follows.

We next prove that $\mathcal{F}_2$ is $\mathrm{VC}$ with $\mathrm{VC}$ index $\leq 3$. Let $p_j=\left(l_j, u_j, c_j\right) \in \mathbb{R}^3$, for $j=1, \ldots, 3$, be three distinct points. Also, define $\mathcal{F}_2^*$ as $\mathcal{F}_2$ but with the permissible range for $t$ expanded to all of $\mathbb{R}$. Then, since $\mathcal{F}_2 \subset \mathcal{F}_2^*$, showing that the new class is $\mathrm{VC}$ with $\mathrm{VC}$ index $\leq 3$ will imply the same for $\mathcal{F}_2$. Define $f_t(l, u)=1_{\{l \leq t<u\}}(t-l)$, for all $(t, l, u) \in \mathbb{R}^3$. Let $T \subset \mathbb{R}$ be eight distinct points. Our task is to show that it is impossible to find such a $T$ such that all 8 possible subsets of the collection $C=\left\{p_1, p_2, p_3\right\}$ can be obtained via sets of the form \begin{equation}\label{sets}
	C \cap\left\{(l, u, c) \in \mathbb{R}^3: f_t(l, u)>c\right\},
\end{equation}

Note that if $c_1<0$, then $p_1$ will never be excluded from sets of the form given in \ref{sets} since $f_{t_1}(l_1,u_1)\geq 0$ for any possible values of $(t_1,l_1,u_1)\in\mathbb{R}^3$. This is also true for $c_2$ and $c_3$. Thus, we will assume going forward without loss of generality that $c_j \geq 0$ for $1 \leq j \leq 3$. Similarly, if $u_1 \leq l_1$, then $f_{t_1}\left(l_1, u_1\right)=0$ for all $t_1 \in \mathbb{R}$. Hence, we will also assume going forward that $l_j<u_j$ for $1 \leq j \leq 3$. Next, let $D_t^*=\left\{(l, u): 1_{\{l<t<u\}}>0\right\}, p_j^*=\left(l_j, u_j\right)$, for $1 \leq j \leq 3$, and $C^*=\left\{p_1^*, p_2^*, p_3^*\right\}$; and note that if we can't obtain all possible eight subsets of $C^*$ from sets of the form $C^* \cap D_t^*$, as $t$ ranges over $T$, then it will be impossible for the sets $D_t$ to shatter $C$. To see this, note first that $f_t(l, u)>0$ if and only if $1_{\{l<t<u\}}>0$, and that $p_1 \in C \cap D_t$ only if $f_t\left(l_1, u_1\right)>0$. Generalizing this, we deduce that $p_j \in C \cap D_t$ only if $p_j^* \in C^* \cap D_t^*, 1 \leq j \leq 3$. The implication does not go in the other way since $c_j$ could be greater than 1 , preventing $p_j$ from being in $C \cap D_t$ even if $p_j^*$ is in $C^* \cap D_t^*$. The conclusion of this is that it is impossible for $\left\{D_t: t \in T\right\}$ to shatter $C$ if $\left\{D_t^*, t \in T\right\}$ does not shatter $C^*$. Thus, if we can show that $\left\{D_t^*: t \in T\right\}$ does not shatter $C^*$, we have established the desired VC index bound.

Let us first try to shatter $\left\{p_1^*, p_2^*\right\}$ with sets of the form $D_t^*$. To do this, we will need the intervals $\left(l_1, u_1\right)$ and $\left(l_2, u_2\right)$ to overlap some but also both have non-zero segments which don't overlap. One way to do this is to have $l_1<l_2<u_1<u_2$. The other equivalent possibility happens when the indices 1 and 2 are swapped. Because of this equivalence, we will just use the initial choice of indices. Now let $T^*=\left\{t_1, \ldots, t_4\right\}$, where $t_1<l_1, l_1<t_2<l_2, l_2<t_3<u_1$, and $u_1<t_4<u_2$. Now it is fairly easy to see that $D_t^*$, as $t$ ranges over $T^*$, shatters $\left\{p_1^*, p_2^*\right\}$. Specifically, $D_t^*$ with $t=t_1$ picks out the null set, $t_2$ picks out $p_1, t_3$ picks out $\left\{p_1, p_2\right\}$, and $t_4$ picks out $p_2$. We can also see that any other non-equivalent arrangement will not result in shattering. Next we need to see if we can add $\left(l_3, u_3\right)$ to this in such a manner that there will exist a $T^*=\left\{t_1, \ldots, t_8\right\}$ such that $D_t^*$, for $t \in T^*$, shatters $C^*$. To do this, we need $\left(l_3, u_3\right)$ to have a portion which does not intersect with either $\left(l_1, u_1\right)$ or $\left(l_2, u_2\right)$, another portion which intersects only with $\left(l_1, u_1\right)$, another that intersects only with $\left(l_2, u_2\right)$, and another portion which intersects with both $\left(l_1, u_1\right)$ and $\left(l_2, u_2\right)$ simultaneously. The impossibility of doing this becomes apparent when visualizing first two intervals above the number line. It is not hard to find an $\left(l_3, u_3\right)$ which overlaps each of the non-null subsets, but once we try to have the nterval reach either below $l_1$ or above $u_2$, we can't do this without encompassing at least one of $\left(l_1, u_1\right)$ or $\left(l_2, u_2\right)$. Thus it is impossible to find any points $\left\{p_1^*, p_2^*, p_3^*\right\}$ and a corresponding $T^*$ which is able to shatter $C^*$. Hence $\mathcal{F}_2$ is VC with VC index $\leq 3$.

We now need to prove that $\mathcal{F}_2$ is pointwise measureable. Let
$$
\mathcal{G}_2=\left\{1_{\{l \leq t<u\}}(t-l): t \in[0,1] \cap \mathbb{Q}\right\},
$$
where $\mathbb{Q}$ is the set of rationals. Fix $t \in[0,1]$, and let $\left\{t_n\right\} \subset[0,1] \cap \mathbb{Q}$ be a sequence such that $t_n \geq t$ for all $n \geq 1$ and $t_n \rightarrow t$, as $n \rightarrow \infty$. Then it is easy to verify that $f_{t_n}(l, u) \rightarrow f_t(l, u)$ for all $(l, u) \in \mathbb{R}^2$. Since $t$ was arbitrary, we have just shown that every function in $\mathcal{F}_2$ is the pointwise limit of a sequence of functions in $\mathcal{G}_2$. Since $\mathcal{G}_2$ is a countable set, we have now verified that $\mathcal{F}_2$ is pointwise measureable. 

Last, every element in $\mathcal{G}$ is bounded between zero and one for all $l,u \in [0,1]$ and therefore $\mathcal{G}$ has a bounded envelope, which is the constant function equal to $1$ for all $l,u \in [0,1]$. The proof is now complete.

\section{Proof of Proposition 1}

If $F$ is continuous on $t$ for all $x$, the statement of the theorem is equivalent to saying:  $\mathbb{P}(F(T,X)\leq t\mid L=l,U=u,X=x)  = 1\{F(l,x)\leq t \leq F(u,x)\}\frac{t-F(l,x)}{F(u,x)- F(l,x)} + 1\{F(u,x) \leq t\}$. In virtue of Assumption 1, it suffices to show the following

$$\begin{aligned}
	&\mathbb{P}(F(T,X) \leq t \mid l \leq T \leq u ,X=x) = \\ &1\{F(l,x)\leq t \leq F(u,x)\}\frac{t- F(l,x)}{F(u,x) - F(l,x)} + 1\{F(u,x) \leq t\}
\end{aligned}$$

This holds because
$$
\begin{aligned}&\mathbb{P}(F(T,X) \leq t \mid l \leq T \leq u ,X=x) =\\& \frac{\mathbb{P}(F(T,x) \leq t , l \leq T \leq u \mid X=x )}{\mathbb{P}(l \leq T\leq u \mid X=x)}= \\& \frac{\mathbb{P}(F(T,x) \leq t , F(l,x) \leq F(T,x) \leq F(u,x) \mid X=x )}{\mathbb{P}(l \leq T\leq u \mid X=x)}= \\ &\frac{\mathbb{P}(F(T,x) \leq t , F(l,x) \leq F(T,x) \leq F(u,x) \mid X=x )}{F(u,x) - F(l,x)}= \\&1\{F(l,x)\leq t \leq F(u,x)\}\frac{t- F(l,x)}{F(u,x) - F(l,x)}+ 1\{F(u,x) \leq t\}
\end{aligned}$$


\section{Proof of Proposition 2}

As split 1 and split 2 are independent, the marginal laws equal the conditional ones.  
Conditional on $\hat F_1$

\begin{flalign*}&\mathbb{P}(\hat F_1(T,X) \leq t \mid l \leq T \leq u ,X=x) = \frac{\mathbb{P}(\hat F_1(T,x) \leq t , l \leq T \leq u \mid X=x )}{\mathbb{P}(l \leq T\leq u \mid X=x)}=\\& \frac{\mathbb{P}(\hat F_1(T,x) \leq t , \hat F_1(l,x) \leq \hat F_1(T,x) \leq \hat F_1(u,x) \mid X=x )}{\mathbb{P}(l \leq T\leq u \mid X=x)}= \\ &\frac{1}{F(u,x) - F(l,x)}1\{\hat F_1(l,x)\leq t \leq  \hat F_1(u,x)\}\mathbb{P}( \hat F_1(l,X)\leq \hat F_1(T,X)\leq t\mid X=x)+\\&+ 1\{ \hat F_1(u,x) \leq t\} 
\end{flalign*}

Define $ F(t, x) = \hat{F}_1(t, x)+ r(t,x)$
Formally, the statement
$$
\sup _{t \leq \tau, x \in \mathbb{R}^p}\left|\hat{S}_1(t, x)-S(t, x)\right|=\mathcal{O}_P\left(r_{N-n}\right)
$$
means that

$$\forall \varepsilon \quad \exists N_{\varepsilon}, \delta_{\varepsilon} \quad \text { such that } P\left(\sup _{t \leq \tau, x \in \mathbb{R}^p}\left|r(t,x)\right| \geq \delta_{\varepsilon}r_N\right) \leq \varepsilon \quad \forall N>N_{\varepsilon}$$

Focusing on the $\mathbb{P}$ term

$$\begin{aligned}&\mathbb{P}\left(\hat{F}_1(l, X) \leq \hat{F}_1(T, X) \leq t \mid X=x\right)=\\ & \mathbb{P}\left(\hat{F}_1(l, X) +  r(T,x) \leq \hat{F}_1(T, X) +  r(T,x) \leq t + r(T,x) \mid X=x\right) =\\ &  \mathbb{P}\left(\hat{F}_1(l, X) +  r(T,x) \leq 
	F(T,x) \leq t + r(T,x) \mid X=x\right)= \\&\mathbb{P}\left(\hat{F}_1(l, X) +  \delta_{\varepsilon}r_N \leq 
	F(T,x) \leq t +  \delta_{\varepsilon}r_N\mid X=x\right) = t - \hat F_1(l,x) \end{aligned}$$

And therefore we have 
$$
\begin{aligned}&\mathbb{P}(\hat F_1(T,X) \leq t \mid l \leq T \leq u ,X=x) =\\ &\frac{1}{F(u,x) - F(l,x)}1\{\hat F_1(l,x)\leq t \leq  \hat F_1(u,x)\}(t-\hat F_1(l,x))+ 1\{ \hat F_1(u,x) \leq t\} 
\end{aligned}$$

The proof is completed using Taylor expansion theorem 

$$\frac{1}{\hat F_1(u,x) - \hat F_1(l,x)} = \frac{1}{ F (u,x) -  F (l,x)} -  \frac{r(l,x)-r(u,x)}{( F(u,x) - F(l,x))^2} + \cdots $$

\section{Proof of Corollary 1}
We need the following auxiliary result to ensure that the interval distribution is an unbiased estimator of \(G\).

\begin{lemma}\label{iter} Let $\Phi$ be uniformly distributed on $[\Lambda,\Upsilon]$ conditional on $\Lambda,\Upsilon$. Then 
	$${E}1\{\Phi \leq \cdot \leq \Upsilon \}= {E}\left[1\{\Lambda \leq \cdot \leq \Upsilon\}\frac{\cdot - \Lambda}{\Upsilon - \Lambda} \right],$$
	
	\noindent where the expectation is taken jointly wrt to $\Phi,\Lambda,\Upsilon$.
\end{lemma}

\begin{proof}Using iterated expectations, LHS is
	
	$${E}{E}\left[1\{\Phi \leq \cdot \leq \Upsilon \}\mid \Lambda, \Upsilon\right]$$
	
	Because of the assumption, we have that the conditional density function of $\Phi$ given $\Lambda, \Upsilon$ is $$g(\phi,\lambda,\upsilon)= \frac{1}{\upsilon - \lambda} 1\{\lambda \leq \phi \leq \upsilon\}$$
	
	We can compute the inner conditional expectation as
	
	$$
	\begin{aligned}
		&E\left[1\{\Phi \leq \cdot \leq \Upsilon \}\mid \Lambda, \Upsilon\right] = E\left[1\{\Phi \leq \cdot \leq \Upsilon \}1\{\Lambda \leq \cdot \leq \Upsilon\}\mid \Lambda, \Upsilon\right]\\ &= \int{1\{\phi \leq \cdot \leq \Upsilon \}1\{\Lambda \leq \cdot \leq \Upsilon\}g(\phi,\Lambda,\Upsilon) d\phi  }\\&
		=1\{\Lambda \leq \cdot \leq \Upsilon\}\int{1\{\phi \leq \cdot \leq \Upsilon \}\frac{1}{\Upsilon - \Lambda} 1\{\Lambda \leq \phi \leq \Upsilon\} d\phi  }  =1\{\Lambda \leq \cdot \leq \Upsilon\}\frac{\cdot - \Lambda}{\Upsilon - \Lambda}
	\end{aligned}$$
	
\end{proof}

Write $G(t)= P(\Phi \leq t) = {E}1\{\Phi \leq t\}$. Fix $t \in [0,1]$. On the one hand, we have $1\{\Phi \leq t\} = 1\{\Phi \leq t \leq \Upsilon \} + 1\{\Upsilon \leq t \}  $ and therefore $G(t)=P(\Phi \leq t) = E1\{\Phi \leq t\} = E1\{\Phi \leq t \leq \Upsilon \} + E1\{\Upsilon \leq t \} $. 

On the other hand, write $g_t(l, u)= 1{\{u \leq t\}}+1{\{l \leq t<u\}}\frac{t-l}{u-l}$ and therefore what we want to show holds iff $E1{\{\Upsilon \leq t\}}+E\left[1{\{\Lambda \leq t<\Upsilon\}}\frac{t-\Lambda}{\Upsilon-\Lambda}\right] =  E1\{\Phi \leq t \leq \Upsilon \} + E1\{\Upsilon \leq t \}$ iff $E\left[1{\{\Lambda \leq t<\Upsilon\}}\frac{t-\Lambda}{\Upsilon-\Lambda}\right] =  E1\{\Phi \leq t \leq \Upsilon \} $, and the last is true in virtue of Lemma 3.4.

\section{Proof of Theorem 3}

The next result constitutes the last auxiliary step to be taken towards deriving the asymptotic guarantees of \texttt{uncervals}. Because of pluging in an estimate $\hat F_1$ instead of using the true $F$ for building the $\Phi_i$'s, the limit of the interval process as the size of $\mathcal{I}_2$ goes to infinity equals $G'=G + b$, with $b = \mathcal{O}_{\mathbf{P}}(r_{N-n})$ in virtue of Proposition 2. 
\begin{lemma}
	Let $G_n$ be the empirical distribution function of $\Phi_1, \ldots, \Phi_n$ conditional on $\widehat{F}_1$ (unknown). Then,
	$$ G_n(t)-\mathbb{I}_n(t)=- b(t) + o_P\left(1\right), \quad t \in [0,1] $$
\end{lemma}

\begin{proof} $G_n(t)-\mathbb{I}_n(t) = G_n(t) - G'(t) + G'(t) -\mathbb{I}_n(t) =  G_n(t) - G(t) -b(t)+ G'(t) -\mathbb{I}_n(t) =-b(t) + o_P\left(1\right) + o_P\left(1\right)= - b(t) + o_P\left(1\right)$ because of the classical Glivenko-Cantelli Theorem and weak convergence of the interval process. 
\end{proof}

We now start te proof of Theorem 3. We denote $P(\cdot)=\mathbb{P}(\cdot \mid \left(X_i, L_i, U_i\right): i \in \mathcal{I}_1)$. Denoting by $V_i=\psi(\Phi_i) = \psi(\hat F_1(T_i, X_i))$ the unobserved scores, 
$$\begin{aligned}
	&P\left(T_{N+1} \in \widehat{\mathcal{C}}_{1-\alpha}^{\text {split,*}}\left(X_{N+1}\right)\mid \mathbb{I}_n\right) = P\left(\psi(\hat{F}_1(T_{N+1}, X_{N+1})) \leq \hat{Q}^*_{\mathcal{I}_2}\mid \mathbb{I}_n\right)\\ &=P\left(V_{N+1}\leq \hat{Q}^*_{\mathcal{I}_2}\mid \mathbb{I}_n\right)  =   P\left(\frac{1}{n} \sum_{i=1}^{n} {1}\left\{V^*_i\geq {V}_{N+1}\right\}>\alpha\mid \mathbb{I}_n\right).  
\end{aligned}
$$

Now for all $v>0$

$$V^*_i\leq v \Longleftrightarrow \left | \Phi^*_i-\frac{1}{2}\right| < v \Longleftrightarrow \Phi^*_i <  \frac{1}{2} + v \textrm{ and } \Phi^*_i > \frac{1}{2} - v. $$

Therefore, 

$${1}\left\{V^*_i\leq v\right\} ={1}\left\{   \Phi^*_i < \frac{1}{2} + v \right\} - {1}\left\{\Phi^*_i < \frac{1}{2} - v\right\}. $$

So that

$$\frac{1}{n} \sum_{i=1}^{n} {1}\left\{V^*_i\leq v\right\} = \mathbb{I}^*_n\left(\frac{1}{2} + v\right)   - \mathbb{I}^*_n \left(\frac{1}{2} - v \right) =:\mathbb{I}^*_n\left(v^+\right)   - \mathbb{I}^*_n \left(v^- \right).$$

Consequently, $$  P\left(v\leq \hat{Q}^*_{\mathcal{I}_2}\mid \mathbb{I}_n\right) = P\left(1+ \mathbb{I}^*_n \left(v^- \right) -\mathbb{I}^*_n\left(v^+\right)   >\alpha\mid \mathbb{I}_n\right). $$ If we add $\mathbb{I}_n\left(v^+\right)$, subtract $1-\mathbb{I}_n\left(v^-\right)$, and multiply by $-\sqrt{n}$ both sides of the inequality we have for $v>0$

\resizebox{1.1\hsize}{!}{
	
	$\begin{aligned}
		& P\left(v\leq \hat{Q}^*_{\mathcal{I}_2}\mid \mathbb{I}_n\right) \\
		&=P\left( \sqrt n\left(\mathbb{I}^*_n\left(v^+\right)-  \mathbb{I}_n \left(v^+ \right)\right) - \sqrt n\left(\mathbb{I}^*_n \left(v^- \right) - \mathbb{I}_n \left(v^- \right) \right) <\sqrt n\left(1-\alpha +\mathbb{I}_n \left(v^- \right)- \mathbb{I}_n \left(v^+ \right)\right) \mid \mathbb{I}_n\right)
		\\
		&=P\left( \sqrt n\left(\mathbb{I}^*_n\left(v^+\right)-  \mathbb{I}_n \left(v^+ \right)\right) <\sqrt n\left(1-\alpha +\mathbb{I}_n \left(v^- \right)- \mathbb{I}_n \left(v^+ \right)\right)  + \sqrt n\left(\mathbb{I}^*_n \left(v^- \right) - \mathbb{I}_n \left(v^- \right) \right) \mid \mathbb{I}_n\right)
		\\
		&\approxtext{Cor. 3.}P\left( \sqrt n\left(\mathbb{I}_n\left(v^+\right)- G' \left(v^+ \right)\right) <\sqrt n\left(1-\alpha +\mathbb{I}_n \left(v^- \right)- \mathbb{I}_n \left(v^+ \right)\right)  + \sqrt n\left(\mathbb{I}^*_n \left(v^- \right) - \mathbb{I}_n \left(v^- \right) \right) \mid \mathbb{I}_n\right) + o_P(1)
		\\
		&=P\left( \sqrt n\left(\mathbb{I}_n\left(v^+\right)- G' \left(v^+ \right)\right) - \sqrt n\left(1-\alpha +\mathbb{I}_n \left(v^- \right)- \mathbb{I}_n \left(v^+ \right)\right)  <  \sqrt n\left(\mathbb{I}^*_n \left(v^- \right) - \mathbb{I}_n \left(v^- \right) \right) \mid \mathbb{I}_n\right) + o_P(1)
		\\
		&\approxtext{Cor. 3}P\left( \sqrt n\left(\mathbb{I}_n\left(v^+\right)- G' \left(v^+ \right)\right) - \sqrt n\left(1-\alpha +\mathbb{I}_n \left(v^- \right)- \mathbb{I}_n \left(v^+ \right)\right)  <  \sqrt n\left(\mathbb{I}_n \left(v^- \right) - G' \left(v^- \right) \right)\right) + o_P(1)
		\\
		&=P\left( \left(\mathbb{I}_n\left(v^+\right)- G' \left(v^+ \right)\right) -\left(\mathbb{I}_n \left(v^- \right) - G' \left(v^- \right) \right) < 1-\alpha +\mathbb{I}_n \left(v^- \right)- \mathbb{I}_n \left(v^+ \right)\right) + o_P(1)
		\\
		&\approxtext{Lemma 2}P\left( o_P(1) < 1-\alpha +\mathbb{I}_n \left(v^- \right)- \mathbb{I}_n \left(v^+ \right) \right) + o_P(1)
		\\
		&=P\left(G_n \left(v^+ \right)-{G}_n \left(v^- \right) + o_P(1) < 1-\alpha +\mathbb{I}_n \left(v^- \right)-{G}_n \left(v^- \right)- \mathbb{I}_n \left(v^+ \right) +{G}_n \left(v^+ \right)\right) + o_P(1)
		\\
		&=P\left(G_n \left(v^+ \right)-{G}_n \left(v^- \right) < 1-\alpha + b(v^-) - b(v^+) + o_P(1)\right) + o_P(1)
		\\
		&=P\left(
		\frac{1}{n} \sum_{i=1}^{n} {1}\left\{{V}_i\leq v\right\}  < 1-\alpha + \mathcal{O}_{\mathbf{P}}(r_{N-n})+ o_P(1)\right) + o_P(1)  \\
	\end{aligned}$
}

Therefore, 
$$\begin{aligned}
	&P\left(V_{N+1}\leq \hat{Q}^*_{\mathcal{I}_2}\mid \mathbb{I}_n\right) = \\&= P\left(
	\frac{1}{n} \sum_{i=1}^{n} {1}\left\{{V}_i\leq V_{N+1}\right\}  < 1-\alpha + \mathcal{O}_{\mathbf{P}}(r_{N-n})+  o_P(1)\right) + o_P(1)\end{aligned}$$. The probability in LHS of the previous line involves randomness just due to resampling and the probability in RHS is conditional on the data indexed by $\mathcal{I}_1$. 
Now by taking expectations over the proper training set on both sides of equal sign we arrive to an equality involving $\mathbb{P}$. 
The rest of the proof goes on as in Theorem 2 in \cite{cherno}.

\section{Proof of Lemma 1}
See the proof for Lemma 1 in \cite{cherno} and use the fact that   $$\begin{aligned}
	n^{-1} \sum_{i=1}^{n} \left(\left|{V}_i-\psi(F_i)\right|\right)^2 &=n^{-1} \sum_{i=1}^{n} \left(\left|\hat{F}_1(T_i,X_i)-\frac{1}{2}\right|-\left|{F}(T_i,X_i)-\frac{1}{2}\right|\right)^2\\&\leq 
	n^{-1} \sum_{i=1}^{n} \left(\left|\hat{F}_1(T_i,X_i)-{F}(T_i,X_i)\right|\right)^2
\end{aligned}.$$

\section{Proof of Corollary 4}

First,
$$
\begin{aligned}
	&\left|{V}_{N+1}-\psi(F_{N+1})\right| = \left|\left|\widehat{F}_1(X_{N+1},T_{N+1}) - \frac{1}{2}\right|-\left|F(X_{N+1},T_{N+1}) - \frac{1}{2}\right|\right| \\ &\leq \left|\widehat{F}_1(X_{N+1},T_{N+1})-F(X_{N+1},T_{N+1})\right| 
\end{aligned}
$$

\noindent Now,
$$\begin{aligned}
	&\left|\widehat{K}({V}_{N+1})-K(\psi(F_{N+1}))\right| =\left|\widehat{K}({V}_{N+1})-{K}({V}_{N+1})\right|+\left|{K}({V}_{N+1})-K(\psi(F_{N+1})) \right|   \\& \leq \mathcal{O}_P\left(\operatorname{max}\{r_{N-n}^{2/3},n^{-\frac{1}{2}}\}\right)+W\left|{V}_{N+1}-\psi(F_{N+1})\right|  =\mathcal{O}_P\left(\operatorname{max}\{r_{N-n}^{2/3},r_{N-n},n^{-1/2}\}\right)
\end{aligned}
$$

\noindent Then, $\left|\widehat{K}({V}_{N+1})-K(\psi(F_{N+1}))\right| = {O}_P\left(\operatorname{max}\{r^{2/3}_{N-n},n^{-1/2}\}\right) $
as $r_{N-n}<r_{N-n}^{2/3}$ for $r_{N-n}<1$. Finally, we have
$$\begin{aligned}
	&P\left(1- \widehat{K}({V}_{N+1})>{O}_P\left(n^{-\frac{1}{2}}\right) +\alpha \right ) + {O}_P\left(n^{-\frac{1}{2}}\right) \\&= P\left(1- {K}(\psi(F_{N+1}))>{O}_P\left(n^{-\frac{1}{2}}\right) + {O}_P\left(\operatorname{max}\{r^{2/3}_{N-n},n^{-1/2}\}\right) +\alpha \right ) + {O}_P\left(n^{-\frac{1}{2}}\right)
	\\&= 1-\alpha + {O}_P\left(\operatorname{max}\{r^{2/3}_{N-n},n^{-1/2}\}\right)
\end{aligned} $$

\section{Statistical literature on uncertainty quantification}

In recent years, uncertainty quantification became an  active research area  \cite{geisser2017predictive, politis2015model}. The impact of uncertainty quantification on data-driven systems has led to a remarkable surge of interest in both applied and theoretical domains. These works delve into the profound implications of uncertainty quantification in statistics and beyond, such as in the biomedical field \cite{Banerji2023}.

Geisser's pioneering book \cite{geisser2017predictive} develops a mathematical theory of predictive inference. Building upon Geisser's foundations, Politis presented a comprehensive methodology \cite{politis2015model} that effectively harnesses resampling techniques. Additionally, the book of Vovk, Gammmerman and Shafer \cite{vovk2005algorithmic} 
has been pivotal, see recent reviews \cite{angelopoulos2021gentle,fontana2023conformal}. 

One of the most widely used and robust frameworks for quantifying uncertainty in statistical and machine learning models is conformal inference \cite{shafer2008tutorial}. The central idea of conformal inference is rooted in the concept of exchangeability \cite{kuchibhotla2020exchangeability}, even though we will assume that the observedd random elements $\mathcal{D}_{N}$ are independent and identically distributed (i.i.d.).
\noindent Now, we present a general overview of conformal inference methods for regresion models with scalar responses. Consider the sequence $\mathcal{D}_{N} = \{(X_i, T_{i})\}^{N}_{i=1}$ of i.i.d. random variables. Given a new i.i.d. random pair $(X, T)$ with respect to $\mathcal{D}_{N}$, conformal prediction provides a family of algorithms for constructing predictive intervals independently of the regression method used, as introduced by \cite{vovk2005algorithmic}.

Fix any regression algorithm 
\[\alg: \ \cup_{N\geq 0} \left(\Xcal\times \R\right)^N \ \rightarrow \
\left\{\textnormal{measurable functions $\widetilde{m}: \Xcal\rightarrow\R$}\right\}, 
\] 
which maps a data set containing any number of pairs $(X_i,T_i)$, to a fitted
regression function $\widetilde{m}$. The algorithm $\alg$ is required to treat
data points symmetrically, i.e.,

\begin{equation}\label{eqn:alg_symmetric}
	\alg\big((x_{\pi(1)},t_{\pi(1)}),\dots,(x_{\pi(N)},t_{\pi(N)})\big) =
	\alg\big((x_1,t_1),\dots,(x_N,t_N)\big)
\end{equation}
for all $N\geq 1$, all permutations $\pi$ on $[N]=\{1,\dots,N\}$, and all
$\{(x_i,t_i)\}_{i=1}^{N}$. Next, for each $t\in\R$, let \[
\widetilde{m}^{t} = \alg\big((X_1,Y_1),\dots,(X_N,T_N),(X,t)\big)
\]
denote the trained model, fitted to the training data together with the test covariate value $X$ and
a hypothesized test response $t$. Let
\begin{equation}\label{eqn:R_y_i}
	R^t_i = 
	\begin{cases}|T_i - \widetilde{m}^{t}(X_i)|, & i=1,\dots,N\\ 
		|t-\widetilde{m}^{t}(X)|, & i=N+1.
	\end{cases}
\end{equation}
The prediction interval for $X$ given by \textit{full conformal} is then defined as 
\begin{equation}\label{eqn:def_fullCP}
	\widehat{C}_{1-\alpha}(X) =\left\{t\in\R \ : \ R^t_{N+1}\leq  (1-\alpha) \textrm{ -quantile of  }\sum_{i=1}^{N+1} \tfrac{1}{N+1} \cdot \delta_{R^t_i}.\right\}
\end{equation}

The \textit{full conformal} method is known to guarantee distribution-free finite-sample predictive coverage at the target level $1-\alpha$: 

\begin{theorem}[Full conformal prediction
	\cite{vovk2005algorithmic}]\label{thm:background_fullCP} 
	If the data points $\mathcal{D}_N \cup \{(X,T)\}$ are i.i.d.\ (or
	more generally, exchangeable), and the algorithm $\alg$ treats the input data
	points symmetrically as in~\eqref{eqn:alg_symmetric}, then the full conformal
	prediction set defined in~\eqref{eqn:def_fullCP} satisfies 
	\[
	\mathbb{P}(T\in \widehat{C}_{1-\alpha}(X)) \geq 1-\alpha.
	\]
\end{theorem}
\noindent The same result holds for split conformal methods, which separates the fitting and ranking steps using sample splitting, and its computational cost is simply that of the fitting step \cite{lei2018distribution}.

Conformal inference techniques have been applied to various regression settings, including estimation of the conditional mean \cite{lei2018distribution}, conditional quantiles \cite{https://doi.org/10.1002/sta4.261}, and different functionals of conditional distributions \cite{singh2023distribution, cherno}. In recent years, multiple extensions of conformal inference techniques have emerged to handle counterfactual inference problems \cite{chernozhukov2021exact, yin2022conformal, jin2023sensitivity}, heterogeneous policy effect \cite{cheng2022conformal}, reinforcement learning \cite{dietterichreinforcement}, federated learning \cite{lu2023federated}, outlier detection \cite{bates2023testing}, hyphotesis testing \cite{hu2023two}, robust optimization \cite{johnstone2021conformal},  multilevel structures \cite{fong2021conformal, dunn2022distribution}, missing data \cite{matabuena2022kernel, zaffran2023conformal}, and survival analysis problems \cite{candesrssb, teng2021t}, as well as problems involving dependent data such as time series and spatial data \cite{chernozhukov2021exact,  xu2021conformal, sun2022conformal, xu2023conformal}.


Asymptotic marginal- and in certain cases, conditional- guarantees for the coverage are provided in  \cite{10.1093/imaiai/iaac017,wu2023bootstrap, das2022model}. The key idea behind these approaches is the application of resampling techniques, originally proposed by  \cite{politis1994large}, to residuals or other score measures. Notably, these methodologies are applicable regardless of the predictive algorithm being used, as emphasized by \cite{politis2015model}.

Bayesian methods are also an important framework to quantify uncertainty (see \cite{chhikara1982prediction}), which can also be integrated with conformal inference methods \cite{angelopoulos2021gentle, fong2021conformal, patel2023variational}. Assuming Gaussian errors and linearity \cite{thombs1990bootstrap, doi:10.1080/00031305.2022.2087735} is a classical and popular approach. However, the latter techniques generally introduce stronger parametric assumptions in statistical modeling and include the limitation or difficulty in selecting the appropriate prior distribution in Bayesian modeling \cite{gawlikowski2023survey}. The theory of tolerance regions gives another connection with the problem studied here \cite{fraser1956tolerance, hamada2004bayesian}, which was generalized for the multivariate case with the notion of depth bands (see  \cite{li2008multivariate}). However, a few conditional depth measures are available in the literature \cite{garcia2023causal}. Depth band measures for statistical objects that take values in metric spaces have recently been proposed \cite{geenens2021statistical, dubey2022depth, liu2022quantiles, virta2023spatial} but only in the unconditional case.


\section{Simulations details}

\subsection{Simulating an interval-censored response}

To simulate interval-censored responses, we modify the function \texttt{simIC\_weib}\footnote{\url{https://github.com/cran/icenReg/blob/master/R/user_utilities.R}} from the \texttt{R} package \texttt{icenReg} \cite{icenreg} so that we can directly provide the regression surface evaluated on the sampled covariates. Consider the following model for the underlying observations: 
\begin{equation}\label{aft}
	p\log\left( sT \right) = -  r(X) + H
\end{equation}

\noindent where $p>0, s>0$, $H$ follows the standard minimum extreme value type I (\textit{Gumbel}) distribution. $r(\cdot)$ is a function of the covariates. The cdf of $T$ conditional on $X=x$ is
$$
\begin{aligned}
	&F(t,x)=\mathbb{P}\left(T \leq t \mid X=x\right)= \mathbb{P}\left(p\log\left( sT \right) \leq p\log\left( st \right)\mid X=x\right) \\& = \mathbb{P}\left(-r(x) + H    \leq  p\log\left( st \right)\right)  \\&= \mathbb{P}\left(H    \leq  p\log\left( st \right) + r(x) \right) 
\end{aligned}
$$

\noindent And therefore we have that 

$$
\begin{aligned}\label{ors}
	&S(t,x)=  1- \mathbb{P}\left(T \leq t \mid X=x\right)\\ &= 1- (1-\exp(-\exp(p\log\left( st \right) + r(x)))) \\& = \exp(-\exp(p\log\left( st \right) + r(x)  ))
\end{aligned}
$$

We use (\ref{ors}) to create a simulation setup (``oracle'') where instead of estimating $\hat F_1$ with a distributional regression algorithm on split 1, we just evaluate $1-S(t,x)$ on split 2. The cumulative hazard is
$$
\begin{aligned}
	H(t,x)= -\log(S(t,x)) = \exp(p\log\left( st \right) +   r(x)) = (st)^p e ^{r(x)}
\end{aligned}
$$

Therefore, Equation \ref{aft} also encompasses a Cox model with baseline cumulative hazard $H_0(t) =(st)^p $, corresponding to a Weibull distribution of scale and shape parameters $s$ and $p$ respectively.  

\subsection{Verification of asymptotic properties}

We prepare four different simulation scenarios, the setup of which is described in Table 1. The underlying true times are simulated according to model \ref{aft}. These are then censored with a case II interval-censoring mechanism \footnote{see \url{https://cran.r-project.org/web/packages/icenReg/index.html}}. Time between inspections is distributed as \texttt{runif(min = 0, max = inspectLength)}. Then, we check the frequency with which the intervals outputted by our algorithm contain the true time-to-event over 100 test points. We average the results across 100 different training sets. This procedure is repeated for six different nominal coverages ($1-\alpha=0.5, 0.7, 0.8, 0.9, 0.95, 0.99$) and four ($N=100,200,500,1000$) different sample sizes. For each row in Table 1, we choose different base estimators for $\hat{F}_1$. In the ``Linear PH'' case, we use \texttt{ic\_sp(Surv(L, U, type = 'interval2'), model = 'ph')} from the \texttt{icenReg} package to fit $\hat{F}_1$. In the ``Non-linear PH'' case, we use Interval Censored Recursive Forests \cite{icrf} with \texttt{ntree=20, nfold=2}. The marginal case is the same as ``Linear PH''  but with no covariates specified. For the ``Oracle'' mode, the true survival function, as derived in \ref{ors}, is directly passed to our approach. 

\begin{table}[ht!]
	\centering
	\begin{tabular}{lccccc}
		\hline
		& $p$ & \texttt{inspections} & \texttt{inspectLength} & $r(X)$ \\ 
		\hline
		Linear PH &2 & 5&0.2 & $-0.1X_1$ \\
		Non-linear PH &2 & 5&0.5 & $\left|5X_2 - 0.5 \right|$\\
		No covariates &0 & 3& 0.5& 0\\
		Oracle & 5 &5 &0.2 &$\sin(\pi X_1) + 2\left|X_2 - 0.5\right| + X_3^3$ \\
		\hline
	\end{tabular}
	\caption{Note the increased difficulty for the ``No covariates'' case, where there are only 3 checkup points. Also, in the ``Linear PH'' case the covariates were simulated with correlation $\rho=0.1$}
	\label{table:your_table_label}
\end{table}

\begin{figure}[t!]
	\centering
	{\includegraphics[width=0.45\linewidth]{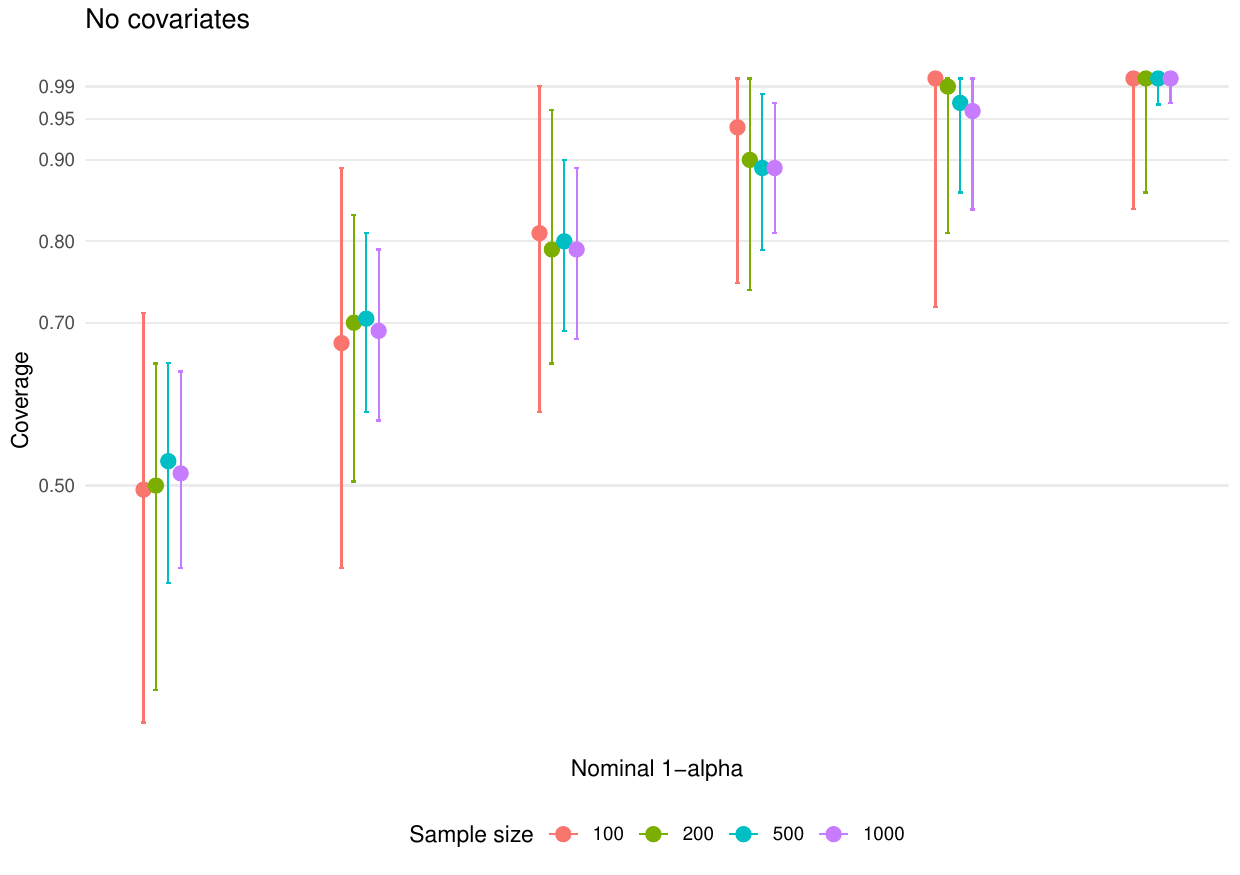}}\quad
	{\includegraphics[width=0.45\linewidth]{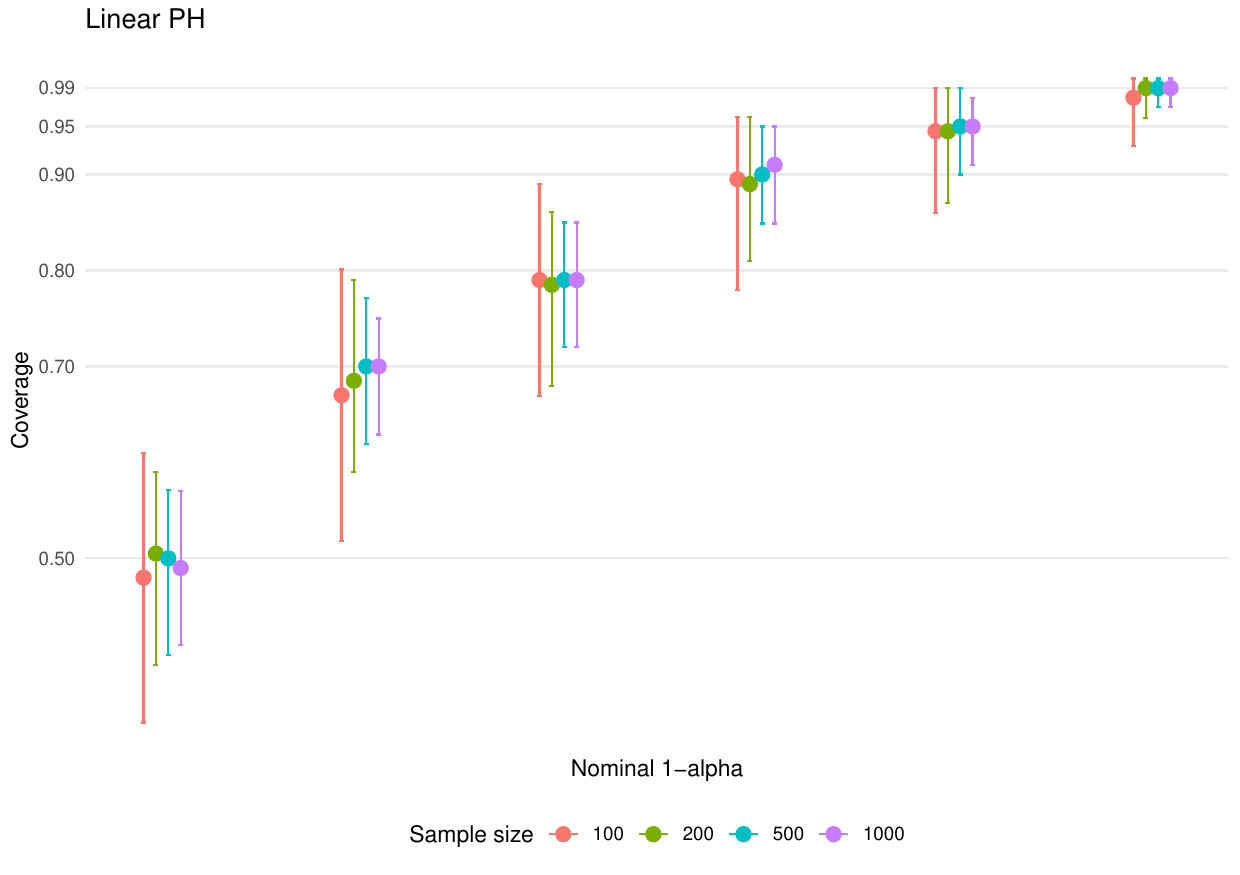}}\\
	{\includegraphics[width=0.45\linewidth]{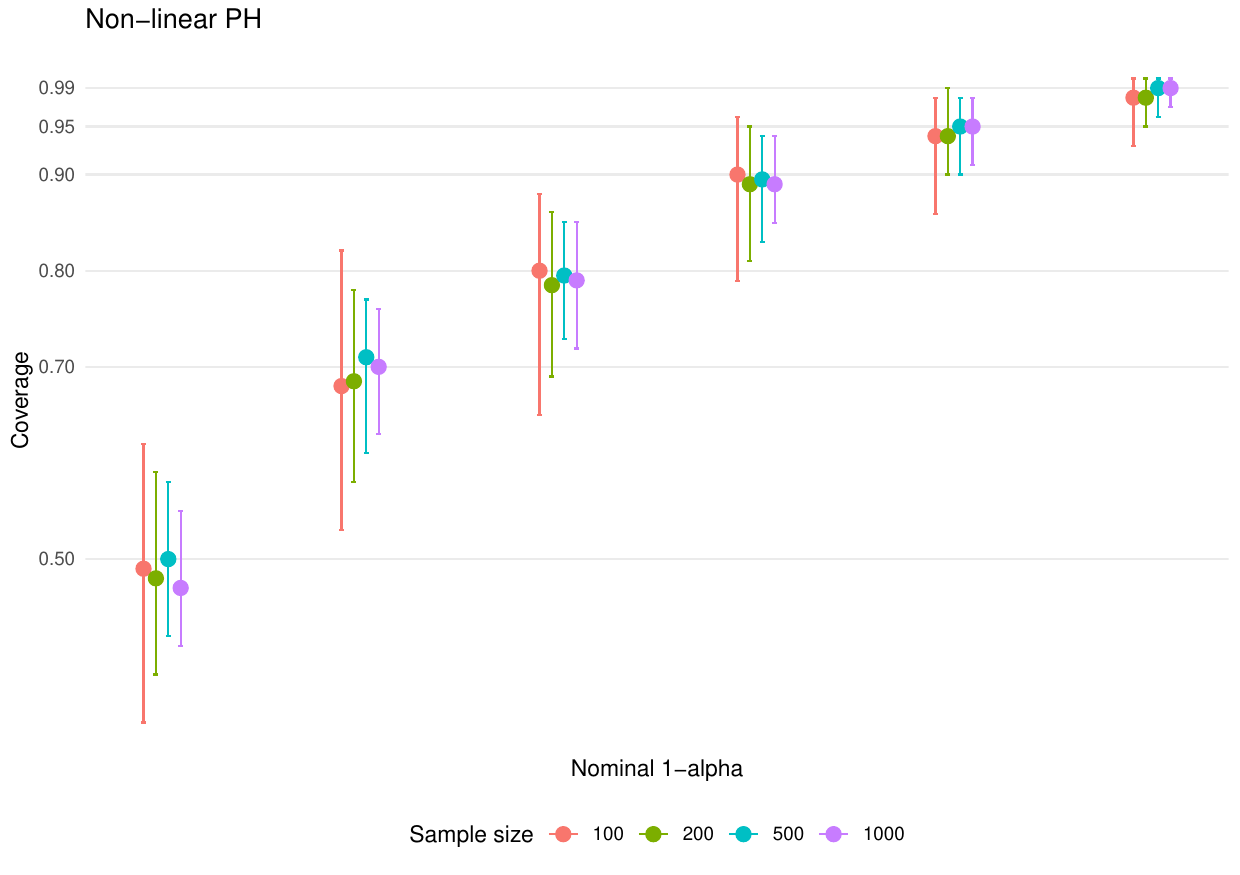}}\quad
	{\includegraphics[width=0.45\linewidth]{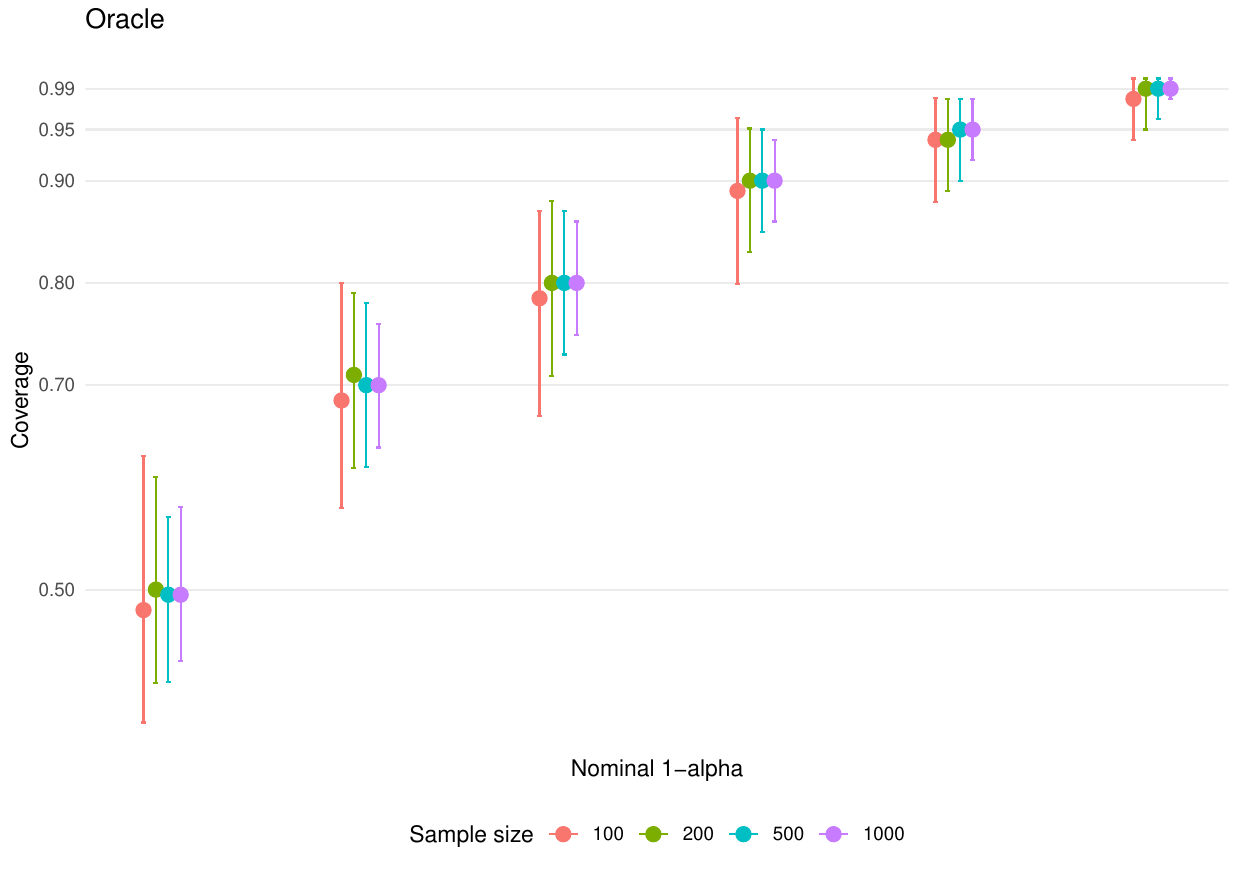}}
	\caption{Averaged over test points vs. nominal coverage for the setup in Section \ref{sim_asym}. The x-axis comprises $\alpha \in \{0.5,0.3,0.2,0.1,0.05,0.01\}$ in all graphs. Sample size refers to $N$. It must be noted that throughout our simulations, \texttt{uncervals} was operating $b=0.5$. This means that we are considering intervals whose image by $\hat F_1(\cdot,x)$ is centered around 0.5.  }
	\label{big_res}
\end{figure}
\newpage
\subsection{Comparing different LPBs}

\begin{table}[h!]
	\centering
	\begin{minipage}{0.18\textwidth}
		\centering
		\caption{ Miscoverage $\cdot 100$ for naive quantile, $\alpha=0.2$}
		\pgfplotstableread[col sep=space]{quantile1.txt}\datatable
		\pgfplotstabletypeset[
		columns/0/.style={column name={$p=1$}},
		columns/1/.style={column name={$p=3$}},
		columns/2/.style={column name={$p=5$}},
		columns/3/.style={column name={$p=10$}},
		every head row/.style={before row=\toprule, after row=\midrule},
		every last row/.style={after row=\bottomrule},
		every row no 0/.style={before row=\midrule}
		]{\datatable}
	\end{minipage}
	\hfill
	\begin{minipage}{0.18\textwidth}
		\centering
		\caption{ Miscoverage $\cdot 100$ for \texttt{ uncervals, e=0} , $\alpha=0.2$}
		\pgfplotstableread[col sep=space]{leftbord1.txt}\datatable
		\pgfplotstabletypeset[
		columns/0/.style={column name={$p=1$}},
		columns/1/.style={column name={$p=3$}},
		columns/2/.style={column name={$p=5$}},
		columns/3/.style={column name={$p=10$}},
		every head row/.style={before row=\toprule, after row=\midrule},
		every last row/.style={after row=\bottomrule},
		every row no 0/.style={before row=\midrule}
		]{\datatable}
	\end{minipage}
	\hfill
	\begin{minipage}{0.18\textwidth}
		\centering
		\caption{ Miscoverage $\cdot 100$ for \texttt{ uncervals, e=*} , $\alpha=0.2$}
		\pgfplotstableread[col sep=space]{alea1.txt}\datatable
		\pgfplotstabletypeset[
		columns/0/.style={column name={$p=1$}},
		columns/1/.style={column name={$p=3$}},
		columns/2/.style={column name={$p=5$}},
		columns/3/.style={column name={$p=10$}},
		every head row/.style={before row=\toprule, after row=\midrule},
		every last row/.style={after row=\bottomrule},
		every row no 0/.style={before row=\midrule}
		]{\datatable}
	\end{minipage}
\end{table}

\begin{table}[h!]
	\centering
	\begin{minipage}{0.18\textwidth}
		\centering
		\caption{ Miscoverage $\cdot 100$ for naive quantile , $\alpha=0.1$}
		\pgfplotstableread[col sep=space]{quantile2.txt}\datatable
		\pgfplotstabletypeset[
		columns/0/.style={column name={$p=1$}},
		columns/1/.style={column name={$p=3$}},
		columns/2/.style={column name={$p=5$}},
		columns/3/.style={column name={$p=10$}},
		every head row/.style={before row=\toprule, after row=\midrule},
		every last row/.style={after row=\bottomrule},
		every row no 0/.style={before row=\midrule}
		]{\datatable}
	\end{minipage}
	\hfill
	\begin{minipage}{0.18\textwidth}
		\centering
		\caption{ Miscoverage $\cdot 100$ for \texttt{ uncervals, e=0}  , $\alpha=0.1$}
		\pgfplotstableread[col sep=space]{leftbord2.txt}\datatable
		\pgfplotstabletypeset[
		columns/0/.style={column name={$p=1$}},
		columns/1/.style={column name={$p=3$}},
		columns/2/.style={column name={$p=5$}},
		columns/3/.style={column name={$p=10$}},
		every head row/.style={before row=\toprule, after row=\midrule},
		every last row/.style={after row=\bottomrule},
		every row no 0/.style={before row=\midrule}
		]{\datatable}
	\end{minipage}
	\hfill
	\begin{minipage}{0.18\textwidth}
		\centering
		\caption{ Miscoverage $\cdot 100$ for \texttt{ uncervals, e=*},   $\alpha=0.1$}
		\pgfplotstableread[col sep=space]{alea2.txt}\datatable
		\pgfplotstabletypeset[
		columns/0/.style={column name={$p=1$}},
		columns/1/.style={column name={$p=3$}},
		columns/2/.style={column name={$p=5$}},
		columns/3/.style={column name={$p=10$}},
		every head row/.style={before row=\toprule, after row=\midrule},
		every last row/.style={after row=\bottomrule},
		every row no 0/.style={before row=\midrule}
		]{\datatable}
	\end{minipage}
\end{table}

\begin{table}[h!]
	\centering
	\begin{minipage}{0.18\textwidth}
		\centering
		\caption{ Miscoverage $\cdot 100$ for naive quantile, $\alpha = 0.05$}
		\pgfplotstableread[col sep=space]{quantile3.txt}\datatable
		\pgfplotstabletypeset[
		columns/0/.style={column name={$p=1$}},
		columns/1/.style={column name={$p=3$}},
		columns/2/.style={column name={$p=5$}},
		columns/3/.style={column name={$p=10$}},
		every head row/.style={before row=\toprule, after row=\midrule},
		every last row/.style={after row=\bottomrule},
		every row no 0/.style={before row=\midrule}
		]{\datatable}
	\end{minipage}
	\hfill
	\begin{minipage}{0.18\textwidth}
		\centering
		\caption{ Miscoverage $\cdot 100$ for \texttt{ uncervals, e=0}, $\alpha=0.05$}
		\pgfplotstableread[col sep=space]{leftbord3.txt}\datatable
		\pgfplotstabletypeset[
		columns/0/.style={column name={$p=1$}},
		columns/1/.style={column name={$p=3$}},
		columns/2/.style={column name={$p=5$}},
		columns/3/.style={column name={$p=10$}},
		every head row/.style={before row=\toprule, after row=\midrule},
		every last row/.style={after row=\bottomrule},
		every row no 0/.style={before row=\midrule}
		]{\datatable}
	\end{minipage}
	\hfill
	\begin{minipage}{0.18\textwidth}
		\centering
		\caption{ Miscoverage $\cdot 100$ for \texttt{ uncervals, e=*},$\alpha=0.05$ }
		\pgfplotstableread[col sep=space]{alea3.txt}\datatable
		\pgfplotstabletypeset[
		columns/0/.style={column name={$p=1$}},
		columns/1/.style={column name={$p=3$}},
		columns/2/.style={column name={$p=5$}},
		columns/3/.style={column name={$p=10$}},
		every head row/.style={before row=\toprule, after row=\midrule},
		every last row/.style={after row=\bottomrule},
		every row no 0/.style={before row=\midrule}
		]{\datatable}
	\end{minipage}
\end{table}
\newpage
\subsection{Conditional coverage}

\begin{figure}[h!]
	\centering
	{\includegraphics[width=0.45\linewidth]{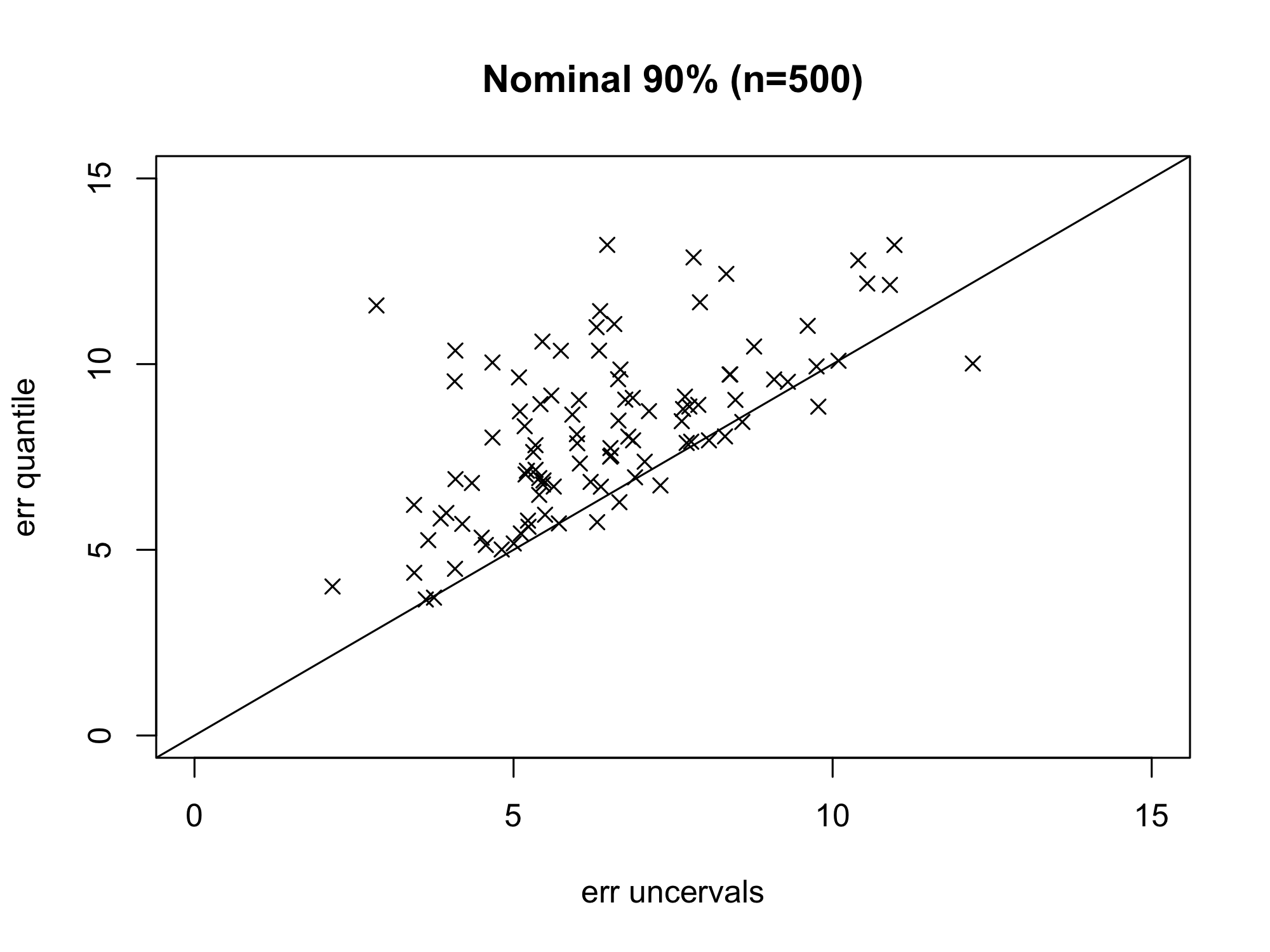}}\quad
	{\includegraphics[width=0.45\linewidth]{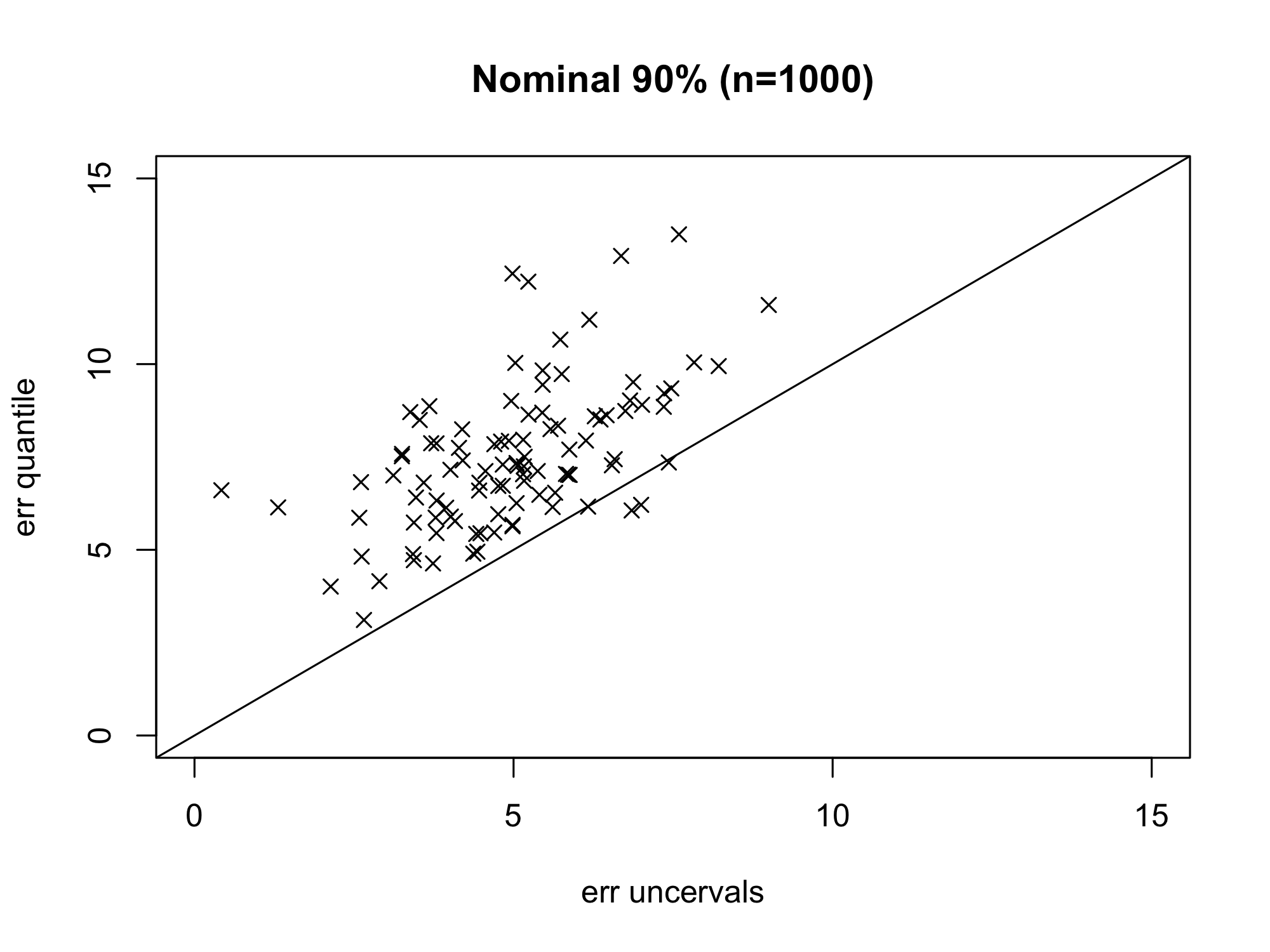}}\\
	
	\caption{ Becnhmark plots for for our approach vs naively taking the $\alpha$- th quantile using the empirical square root $L_2$ error between the estimated coverage propensity and nominal $(1-\alpha)$ (\texttt{err}). Notice how $\texttt{uncervals}$ performs better most of the times, especially when increasing the sample size to $n=1000$. See plots for $\alpha=0.1$ in the supplement }
	\label{fig:best2}
\end{figure}

\begin{figure}[h!]
	\centering
	{\includegraphics[width=0.45\linewidth]{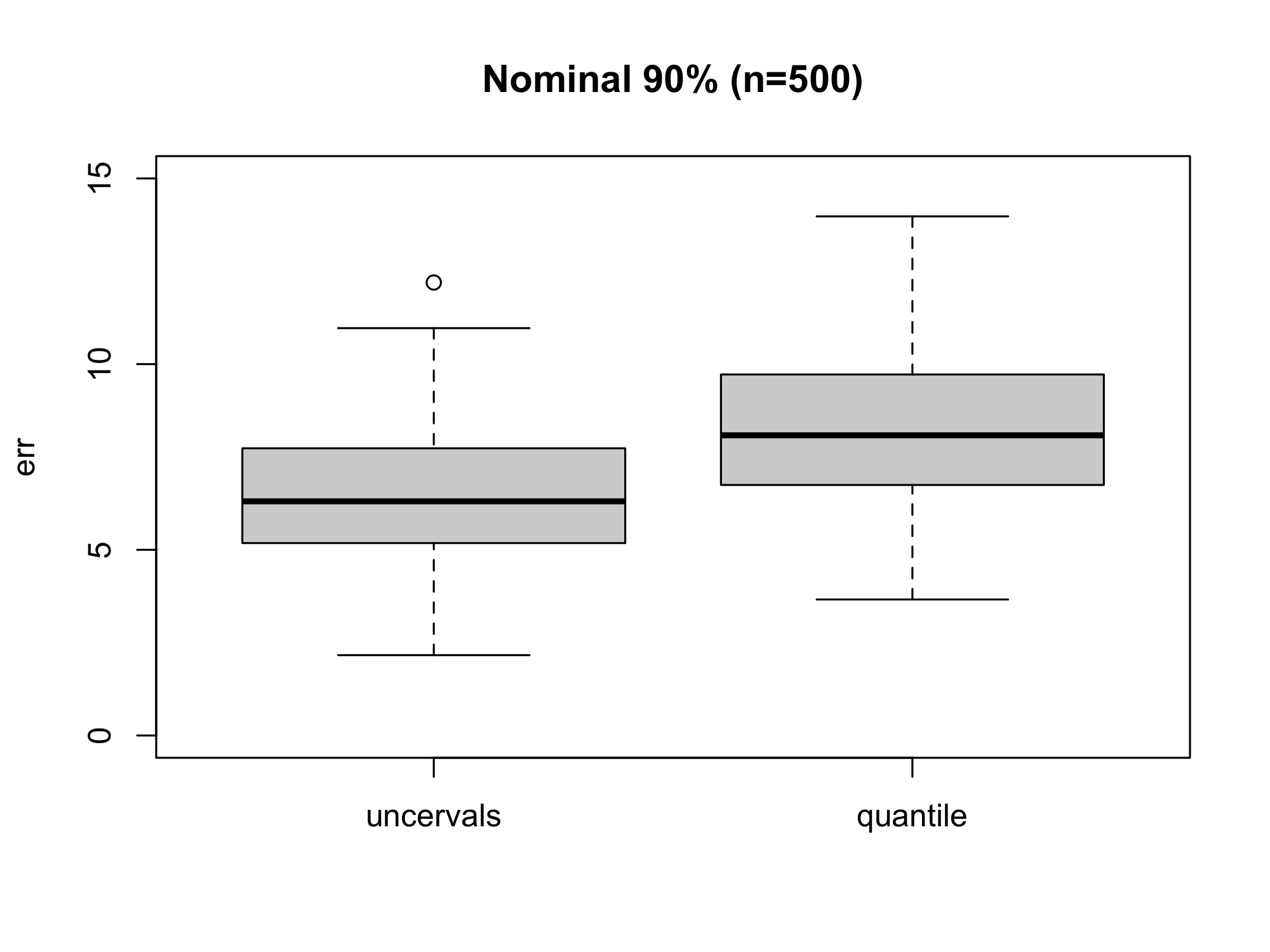}}\quad
	{\includegraphics[width=0.45\linewidth]{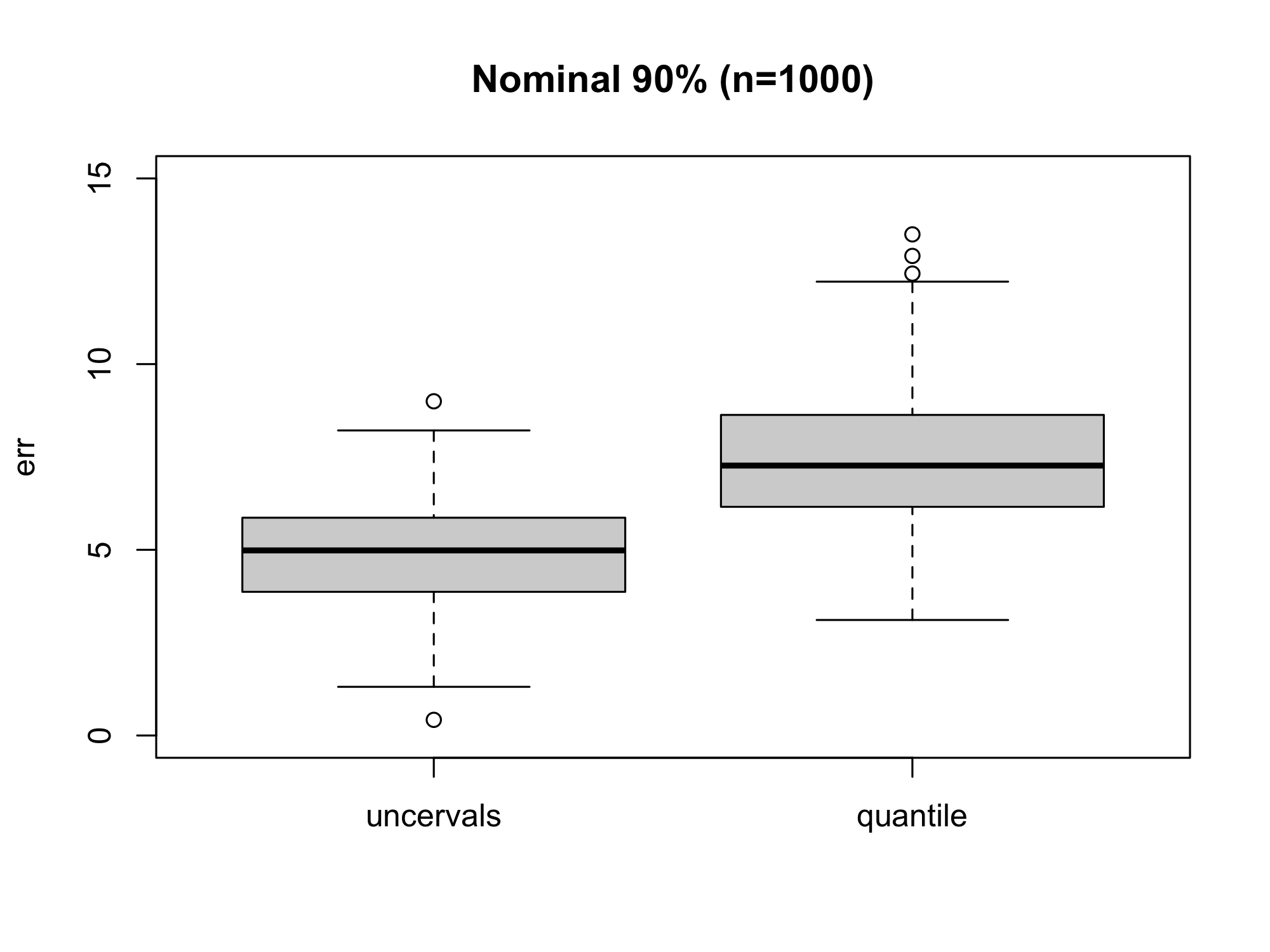}}\\
	{\includegraphics[width=0.45\linewidth]{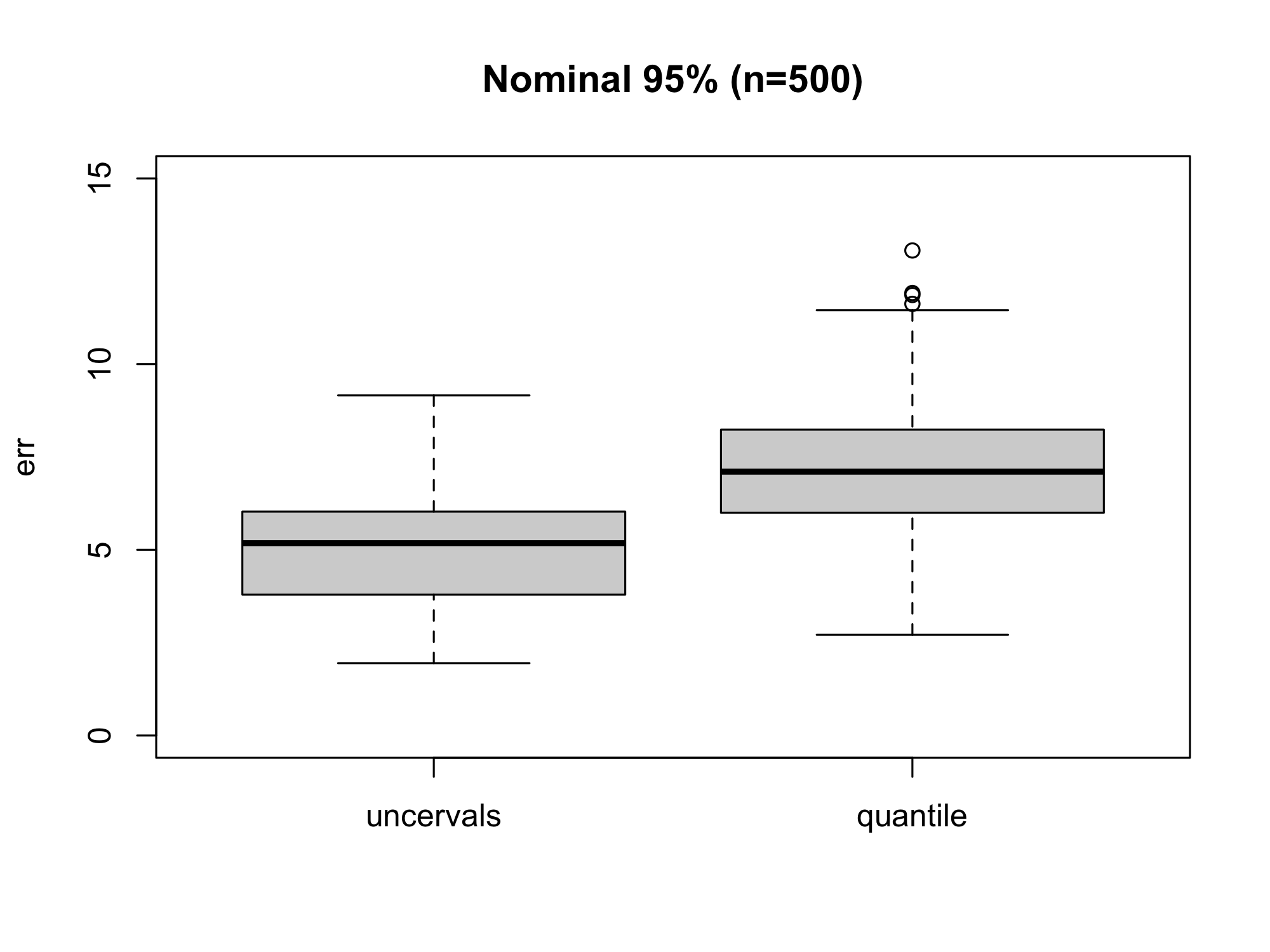}}\quad
	{\includegraphics[width=0.45\linewidth]{100095.png}}
	\caption{Boxplots for the results condensated in the tables in the section of the main document with the same title.  }
	\label{fig:boxs}
\end{figure}
\end{document}